\documentclass[twocolumn]{svjour3}  
\smartqed  
\RequirePackage[pdftex]{graphicx}
\RequirePackage{mathptmx}      
\usepackage[utf8]{inputenc} 
\usepackage{amsmath,amssymb}
\usepackage[unicode,hypertexnames,setpagesize,%
    pdftex,%
    colorlinks,%
    citecolor=blue,%
    hyperindex,%
    plainpages=false,%
    bookmarksopen,%
    bookmarksnumbered%
  ]{hyperref}

\providecommand\texorpdfstring[2]{#1}
\usepackage[numbers,square,comma,sort&compress]{natbib}

\usepackage{textcomp}

\usepackage{lineno,xcolor}
\def\kt{\ensuremath{k_t}}
\newcommand{\Pmax}{p}
\newcommand{\CCFM}{CCFMa,CCFMb,Catani:1989sg,CCFMd}

\usepackage{tikz}
\usetikzlibrary{arrows,shapes,positioning}

\usepackage{multirow}

\usepackage{xspace}
\providecommand{\fitter}{{\texttt{HERA\-Fitter}}\xspace}
\providecommand{\fastnlo}{{\texttt{fastNLO}}\xspace}
\providecommand{\applgrid}{{\texttt{APPL\-GRID}}\xspace}
\providecommand{\qcdnum}{{\texttt{QCDNUM}}\xspace}
\providecommand{\minuit}{{\texttt{MINUIT}}\xspace}
\providecommand{\mcfm}{{\texttt{MCFM}}\xspace}
\providecommand{\difftop}{{\texttt{DiffTop}}\xspace}
\providecommand{\nlojetpp}{{\texttt{NLOJet++}}\xspace}
\providecommand{\lhapdf}{{\texttt{LHAPDF}}\xspace}

\providecommand{\GeV}{\ensuremath{\,\text{Ge\hspace{-.08em}V}}\xspace}
\providecommand{\pperp}{\ensuremath{p_{\perp}}\xspace}
\providecommand{\mur}{\ensuremath{\mu_\mathrm{R}}\xspace}
\providecommand{\muf}{\ensuremath{\mu_\mathrm{F}}\xspace}
\providecommand{\as}{\ensuremath{\alpha_\mathrm{s}}\xspace}
\providecommand{\asmz}{\ensuremath{\alpha_\mathrm{s}(m_Z)}\xspace}

\providecommand{\tmdlib}{{\texttt{TMDlib}}\xspace}

\journalname{DESY Report 14-188}
\begin{document}\sloppy

\title{HERAFitter 
}
\subtitle{Open Source QCD Fit Project \\  }


\author{
S.~Alekhin$^{1,2}$\and
O.~Behnke$^{3}$\and
P.~Belov$^{3,4}$\and
S.~Borroni$^{3}$\and
M.~Botje$^{5}$\and
D.~Britzger$^{3}$\and
S.~Camarda$^{3}$\and 
A.M.~Cooper-Sarkar$^{6}$\and 
K.~Daum$^{7,8}$\and
C.~Diaconu$^{9}$\and 
J.~Feltesse$^{10}$\and
A.~Gizhko$^{3}$\and
A.~Glazov$^{3}$\and
A.~Guffanti$^{11}$\and
M.~Guzzi$^{3}$\and
F.~Hautmann$^{12,13,14}$\and
A.~Jung$^{15}$\and
H.~Jung$^{3,16}$\and
V.~Kolesnikov$^{17}$\and    
H.~Kowalski$^{3}$\and
O.~Kuprash$^{3}$\and
A.~Kusina$^{18}$\and
S.~Levonian$^{3}$\and
K.~Lipka$^{3}$\and
B.~Lobodzinski$^{19}$\and 
K.~Lohwasser$^{1,3}$\and
A.~Luszczak$^{20}$\and    
B.~Malaescu$^{21}$\and
R.~McNulty$^{22}$\and
V.~Myronenko$^{3}$\and
S.~Naumann-Emme$^{3}$\and
K.~Nowak$^{3,6}$\and
F.~Olness$^{18}$\and 
E.~Perez$^{23}$\and
H.~Pirumov$^{3}$\and
R.~Pla\v cakyt\. e$^{3}$\and
K.~Rabbertz$^{24}$\and    
V.~Radescu$^{3}$\and
R.~Sadykov$^{17}$\and
G.P.~Salam$^{25,26}$\and
A.~Sapronov$^{17}$\and
A.~Sch\"oning$^{27}$\and
T.~Sch\"orner-Sadenius$^{3}$\and
S.~Shushkevich$^{3}$\and    
W.~Slominski$^{28}$\and    
H.~Spiesberger$^{29}$\and
P.~Starovoitov$^{3}$\and    
M.~Sutton$^{30}$\and    
J.~Tomaszewska$^{31}$\and    
O.~Turkot$^{3}$\and
A.~Vargas$^{3}$\and
G.~Watt$^{32}$\and 
K.~Wichmann$^{3}$
}
\institute{$ $ Deutsches Elektronen-Synchrotron (DESY), Platanenallee 6, D–15738 Zeuthen, Germany \\
 $ ^{2}$ Institute for High Energy Physics,142281 Protvino, Moscow region, Russia \\
 $ ^{3}$ Deutsches Elektronen-Synchrotron (DESY), Hamburg, Germany\\
 $ ^{4}$ Current address: Department of Physics, St. Petersburg State University, Ulyanovskaya 1, 198504 St. Petersburg, Russia\\
 $ ^{5}$ Nikhef, Science Park, Amsterdam, the Netherlands \\
 $ ^{6}$ Department of Physics, University of Oxford, Oxford, United Kingdom \\
 $ ^{7}$ Fachbereich C, Universit\"at Wuppertal, Wuppertal, Germany \\
 $ ^{8}$ Rechenzentrum, Universit\"at Wuppertal, Wuppertal, Germany \\
 $ ^{9}$ Aix Marseille Universite, CNRS/IN2P3, CPPM UMR 7346, 13288 Marseille, France\\
 $ ^{10}$ CEA, DSM/Irfu, CE-Saclay, Gif-sur-Yvette, France \\
 $ ^{11}$ Niels Bohr International Academy and Discovery Center, Niels Bohr Institute, University of Copenhagen, \\
 $ ^{}$   \hspace{0.15cm} Blegdamsvej 17, DK-2100 Copenhagen, Denmark \\
 $ ^{12}$ School of Physics and Astronomy, University of Southampton, UK \\
 $ ^{13}$ Rutherford Appleton Laboratory, Chilton OX11 0QX, United Kingdom \\
 $ ^{14}$ Dept. of Theoretical Physics, University of Oxford, Oxford OX1 3NP, United Kingdom \\
 $ ^{15}$ FERMILAB, Batavia, IL, 60510, USA \\
 $ ^{16}$ Elementaire Deeltjes Fysica, Universiteit Antwerpen, B 2020 Antwerpen, Belgium  \\
 $ ^{17}$ Joint Institute for Nuclear Research (JINR), Joliot-Curie 6, 141980, Dubna, Moscow Region, Russia \\
 $ ^{18}$ Southern Methodist University, Dallas, Texas \\
 $ ^{19}$ Max Planck Institut F\"ur Physik, Werner Heisenberg Institut, F\"ohringer Ring 6, Mu\"nchen \\
 $ ^{20}$ T. Kosciuszko Cracow University of Technology \\
 $ ^{21}$ Laboratoire de Physique Nucl\' eaire et de Hautes Energies, UPMC and Universit\'e, Paris-Diderot and CNRS/IN2P3, Paris, France \\
 $ ^{22}$ University College Dublin, Dublin 4, Ireland \\
 $ ^{23}$ CERN, European Organization for Nuclear Research, Geneva, Switzerland \\
 $ ^{24}$ Institut f\" ur Experimentelle Kernphysik, Karlsruhe, Germany \\
 $ ^{25}$ CERN, PH-TH, CH-1211 Geneva 23, Switzerland \\
 $ ^{26}$ leave from LPTHE; CNRS UMR 7589; UPMC Univ. Paris 6; Paris 75252, France \\
 $ ^{27}$ Physikalisches Institut, Universit\"at Heidelberg, Heidelberg, Germany \\
 $ ^{28}$ Jagiellonian University, Institute of Physics, Reymonta 4, PL-30-059 Cracow, Poland \\
 $ ^{29}$ PRISMA Cluster of Excellence, Institut f\"ur Physik (WA THEP), Johannes-Gutenberg-Universit\" at, D-55099 Mainz, Germany \\
 $ ^{30}$ University of Sussex, Department of Physics and Astronomy, Sussex House, Brighton BN1 9RH, United Kingdom \\
 $ ^{31}$ Warsaw University of Technology, Faculty of Physics, Koszykowa 75, 00-662 Warsaw, Poland \\
 $ ^{32}$ Institute for Particle Physics Phenomenology, Durham University, Durham, DH1 3LE, United Kingdom \\
}
%

%
%



\date{}

\setcounter{tocdepth}{4}
\maketitle

\begin{abstract}
\fitter is an open-source package
that provides a framework for the determination of the
parton distribution functions (PDFs) of the proton and for
many different kinds of analyses in Quantum Chromodynamics (QCD).
It encodes results from a wide range of experimental measurements in lepton-proton deep inelastic scattering and proton-proton (proton-antiproton) collisions at hadron colliders.
These are complemented with a variety of theoretical options for calculating PDF-dependent cross section predictions corresponding to the measurements.
The framework covers a large number of the existing
methods and schemes used for PDF determination.
%
The data and theoretical predictions are brought together through numerous methodological options for carrying out PDF fits and plotting tools to help visualise the results. 
While primarily based on the approach of collinear factorisation, \fitter also provides facilities for fits of dipole models and transverse-momentum dependent PDFs.
The package can be used to study the impact of new precise measurements from hadron colliders.
This paper describes the general structure of \fitter and its wide choice of options.  
\end{abstract}


\section{Introduction}
\label{sec:intro}
The recent discovery of the Higgs boson \cite{Aad:2012tfa,Chatrchyan:2012ufa} 
and the extensive searches
for signals of new physics in LHC proton-proton collisions
demand high-precision calculations to test the validity of the Standard Model (SM)
and factorisation in Quantum Chromodynamics (QCD).
Using collinear factorisation, inclusive cross sections in hadron collisions may be written as
%
\begin{eqnarray}
\small
%
\sigma(\as(\mur^2),\mur^2,\muf^2)&=& \sum\limits_{a,b}\,  \int\limits_{0}\limits^{1} dx_1\ dx_2  f_a(x_1,\muf^2) f_b(x_2,\muf^2) \nonumber \\ 
&\times&  \, \hat{\sigma}^{ab}(x_1,x_2;\as(\mur^2),\mur^2,\muf^2)  \nonumber \\
&+&{\cal O}\left(\frac{\Lambda_{QCD}^2}{Q^2}\right)\,
\label{eq:fact}
\end{eqnarray}
where the cross section $\sigma$
is expressed
as a convolution of Parton Distribution Functions (PDFs) $f_a$ and $f_b$
with the parton cross section
$\hat{\sigma}^{ab}$,  involving a momentum transfer 
$q$ such that $Q^2 = |q^2| \gg \Lambda_{QCD}^2$, where $\Lambda_{QCD}$ is the QCD scale.  
At Leading-Order (LO) in the perturbative expansion of the strong-coupling constant, the PDFs represent 
the probability of finding a specific parton $a$ ($b$) in the first (second) hadron carrying a fraction $x_1$ ($x_2$) of its momentum.
The indices $a$ and $b$ in Eq.~\ref{eq:fact} indicate the various 
kinds of partons,
i.e. gluons, quarks and antiquarks of different flavours
that are considered
as the constituents of the proton.
The PDFs depend on the factorisation scale, $\muf$, while the parton cross sections depend on the strong coupling constant,
$\as$, and the factorisation and renormalisation scales,
$\muf$ and $\mur$.
The parton cross sections $\hat\sigma^{ab}$ are calculable in perturbative QCD (pQCD) whereas
PDFs are usually constrained by global fits to a variety of experimental data. The assumption that PDFs are universal, within a particular factorisation scheme \cite{Collins:1981uw,Collins:1983ju,Collins:1985ue,Collins:1989gx,Collins:2011zzd}, is crucial to this procedure.
Recent review articles on PDFs can be found in Refs. \cite{Perez:2012um,Forte:2013wc}. 
%
%

A precise determination of PDFs as a function of $x$ requires large amounts of
experimental data that cover a wide kinematic region and that are sensitive to different kinds of partons. Measurements of inclusive Neutral Current (NC) and Charge Current (CC) Deep Inelastic Scattering (DIS) at the lepton-proton ($ep$) collider HERA provide crucial information for determining the PDFs. 
The low-energy fixed target data and different processes from proton-proton ($pp$) collisions at the LHC 
and proton-antiproton ($p \bar p$) collisions at the Tevatron provide complementary information to the HERA DIS measurements.
 The PDFs are determined
from $\chi^2$ fits of the theoretical predictions to the 
data. 
The rapid flow of new data from the LHC experiments and the corresponding theoretical developments, which are providing predictions for more complex processes at increasingly higher orders, has motivated the development of a tool to combine them  together in a fast, efficient, open-source framework.
%

This paper describes the open-source QCD fit framework \fitter~\cite{herafitter:page}, which includes a set of tools to facilitate global 
QCD analyses of $pp$, $p\bar{p}$ and $ep$ scattering data. 
It has been developed for the determination of PDFs and the extraction of fundamental parameters of QCD such as the heavy
quark masses and the strong coupling constant. It also provides a common framework for the
comparison of different theoretical approaches. Furthermore, it can be used to test the impact 
of new experimental data on the PDFs and on the SM parameters.

This paper is organised as follows:
The general structure of \fitter is presented in Sec.~\ref{sec:structure}.
In Sec.~\ref{sec:theory} the various processes available in \fitter
and the corresponding theoretical calculations, performed within the framework of collinear factorisation and the DGLAP~\cite{Gribov:1972ri,Gribov:1972rt,Lipatov:1974qm,
Dokshitzer:1977sg,Altarelli:1977zs} formalism, are discussed. In
Sec.~\ref{sec:techniques} tools for fast calculations of the theoretical predictions are presented.
In Sec.~\ref{sec:method} the 
methodology to determine PDFs through fits based on various
 $\chi^2$ definitions is described. In particular, different treatments of correlated experimental uncertainties are presented.
Alternative approaches to the DGLAP formalism are presented in Sec.~\ref{sec:alternative}.
The organisation of the \fitter code is discussed in Sec. \ref{sec:organisation}, specific applications 
of the package are presented in Sec.~\ref{sec:examples}, which is followed by a summary in Sec.~\ref{sec:summary}.
%

\section{The HERAFitter Structure}
\label{sec:structure}


The diagram in Fig. \ref{fig:flow} gives a schematic overview
of the \fitter structure and functionality, which can be divided into four main blocks:
\begin{figure}[!ht]
  \begin{tikzpicture}[node distance=1cm, auto,>=latex', thick]
      \path[->] node[draw, text width=2cm, text centered] at (0,0) (init) {\bf Initialisation};
      \path[->] node[draw, below left=0.3cm and -0.7cm of init, text width=3.2cm] (data) 
                    {\begin{center} \vspace{-0.3cm}{\bf Data} 
		     \end{center} 
		     {\scriptsize 
		     \begin{itemize}
                      \vspace{-0.3cm}
		      \item Collider, Fixed Target: $ep$, $\mu p$
		      \item Collider: $pp, p\bar p$
		     \end{itemize}}
		     } (init) edge (data);
      \path[->] node[draw, below right=0.3cm and -0.7cm of init, text width=3.55cm] (theory) 
                    {\begin{center} \vspace{-0.3cm}{\bf Theory} 
		     \end{center} 
		     {\scriptsize 
		     \begin{itemize}
                      \vspace{-0.3cm}
		      \item PDF Parametrisation
              \item QCD Evolution:  \\
                   DGLAP (\qcdnum), non-DGLAP (CCFM, dipole)
		      \item Cross Section Calculation
		     \end{itemize}}
		     } (init) edge (theory);
      \path[->] node[draw, below right=1.1cm and -1.7cm of data, text width=4cm] (minuit) 
                    {\begin{center} \vspace{-0.3cm}{\bf QCD Analysis} 
                      \vspace{-0.2cm}
		     \end{center} 
		     {\scriptsize 
                      \vspace{-0.1cm}
		     \begin{itemize}
		     \item Treatment of the Uncertainties
             \item Fast $\chi^2$ Computation 
             \item Minimisation (\minuit)
		     \end{itemize}}
		     } (data) edge (minuit)
		     (theory) edge (minuit)
		     (data) ++ (1.45,0) edge [<->,double equal sign distance] ++(1.36,0) (theory);
      \path[->] node[draw, below =0.4cm  of minuit, text width=3.7cm] (res) 
                    {\begin{center} \vspace{-0.3cm}{\bf Results} 
		     \end{center} 
		     {\scriptsize 
		     \begin{itemize}
                      \vspace{-0.3cm}
		      \item PDFs, \lhapdf, \tt TMDlib \rm Grids
		      \item \as, $m_C$, \dots
		      \item Data vs. Predictions
		      \item \(\chi^2\), Pulls, Shifts
		     \end{itemize}}
		     } (minuit) edge (res);
  \end{tikzpicture}
  \caption{Schematic overview of the \fitter program.} 
  \label{fig:flow}
\end{figure}
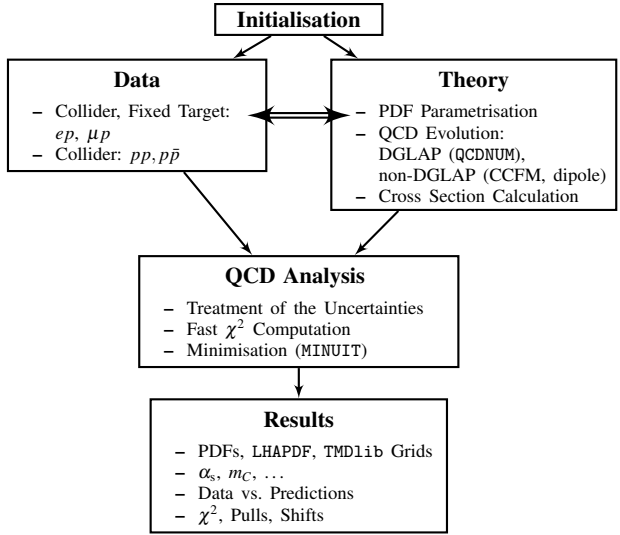

\paragraph{Data:} Measurements from various processes
are provided in the \fitter package including the information on their uncorrelated 
and correlated uncertainties. HERA inclusive scattering data 
are directly sensitive to quark PDFs and indirectly sensitive to the gluon PDF through scaling violations
and the longitudinal structure function $F_L$.  
These data are the basis of any proton PDF extraction, and are used in all current PDF sets 
from MSTW \cite{MSTWpdf}, CT \cite{CT10pdf}, NNPDF \cite{NNPDFpdf}, ABM \cite{Alekhin:2013nda}, JR \cite{Jimenez-Delgado:2014twa} 
and HERAPDF \cite{h1zeus:2009wt} groups. Measurements of charm and beauty quark production at HERA are 
sensitive to heavy quark PDFs and jet measurements have direct sensitivity to the gluon PDF. However, the kinematic range 
of HERA data mostly covers low and medium ranges in $x$.
Measurements from the fixed target experiments, the Tevatron and the LHC provide additional constraints
on the gluon and quark distributions at high-$x$, better understanding of heavy quark 
distributions and decomposition of the light-quark sea. For these purposes, measurements from fixed-target
experiments, the Tevatron and the LHC are included. 

%
%
The processes that are currently available within the \fitter framework are listed in Tab.~\ref{tab:proc}.

\begin{figure}[!ht]
   \centering
   \includegraphics[width=8cm]{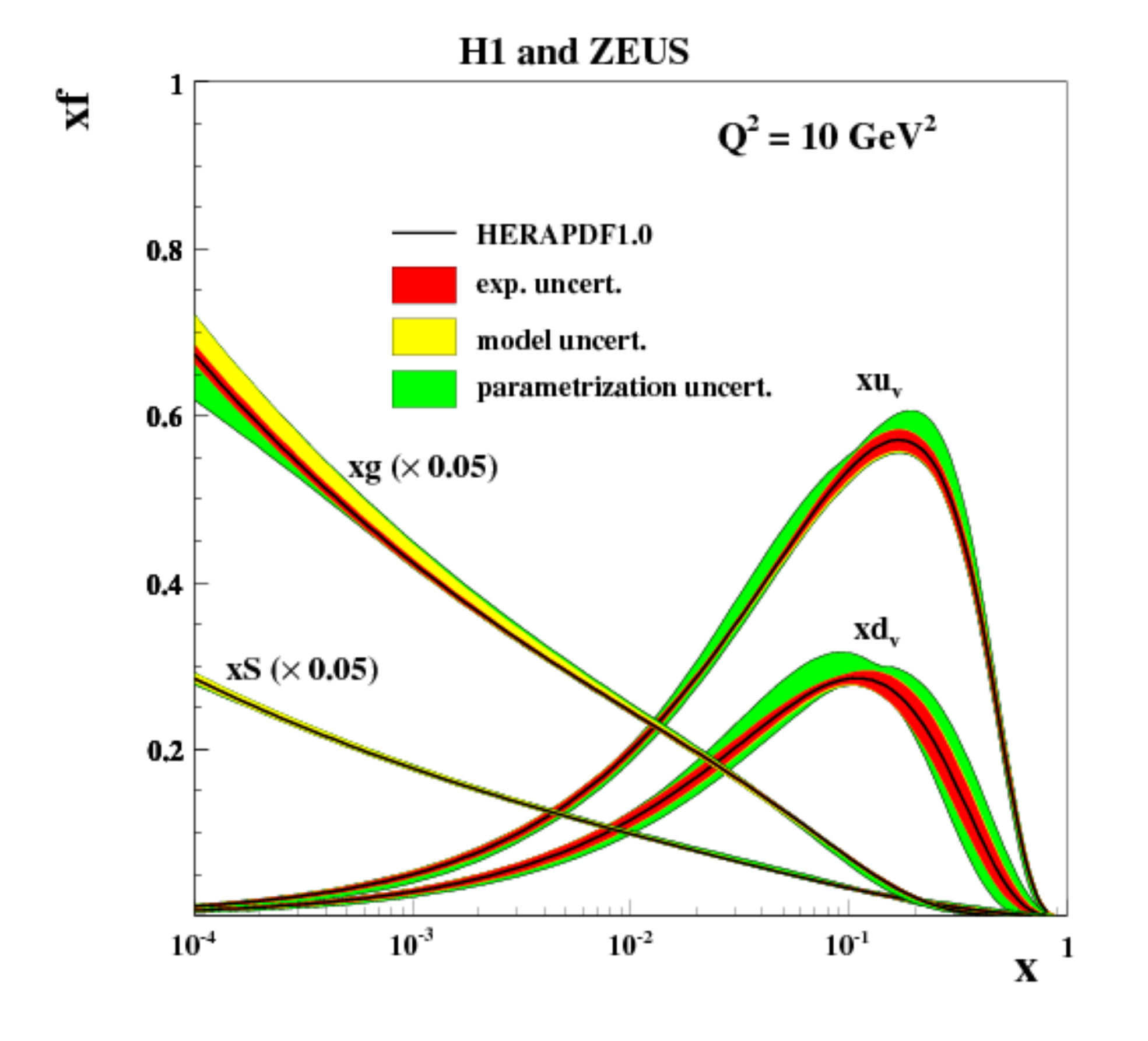}
   \caption{Distributions of valence ($xu_v$, $xd_v$), sea ($xS$) and the gluon ($xg$) PDFs in HERAPDF1.0~\cite{h1zeus:2009wt}. 
       The gluon and the sea distributions are scaled down by a factor of 20.
       The experimental, model and parametrisation uncertainties are shown as coloured bands.}
 \label{fig:hera1}
\end{figure}

\begin{table}
\small
\scriptsize

\begin{tabular}{|l|l|l|l|}
\hline 
\textbf{Experimental} &\textbf{Process}&\textbf{Reaction}&\textbf{Theory schemes} \\
\textbf{Data}         &        &                &\textbf{calculations}  \\
\hline \hline \\ [-2.5ex]
HERA, &DIS NC   &$ep\to eX$      & TR$^\prime$, ACOT, \\
Fixed Target   &  &$\mu p \to\mu X$                 & ZM (\qcdnum), \\
     &         &                & FFN (\texttt{OPENQCDRAD}, \\
     &         &                & \qcdnum), \\ 
     &         &                & TMD (uPDFevolv) \\ [0.5ex]
\hline \\ [-2.5ex]
HERA &DIS CC   &$ep\to \nu_e X$ & ACOT, ZM (\qcdnum), \\
     &         &                & FFN (\texttt{OPENQCDRAD)} \\  [0.5ex]
\cline{2-4}  \\ [-2.0ex]
     &DIS jets &$ep\to e\ \mathrm{jets}X$      & \nlojetpp (\fastnlo)\\ [0.5ex]
\cline{2-4} \\ [-2.0ex]
     &DIS heavy & $ep\to e c \bar{c} X$, &   TR$^\prime$, ACOT, \\
     &quarks         & $ep\to e b \bar{b} X$ & ZM (\qcdnum), \\
     &         &                & FFN (\texttt{OPENQCDRAD}, \\
     &         &                & \qcdnum) \\  [0.5ex]
\hline \\ [-2.5ex]
Tevatron,&Drell-Yan &$pp(\bar p)\to l\bar l X$, & \mcfm (\applgrid) \\
LHC              &          &$pp(\bar p)\to l\nu  X$ &                 \\ [0.5ex]
\cline{2-4}  \\ [-2.0ex]
              &top pair   &$pp(\bar p) \to t\bar t X$  & \mcfm (\applgrid),  \\
              &            &                            & \texttt{HATHOR}, \difftop \\ [0.5ex] 
\cline{2-4}  \\ [-2.0ex]
              &single top &$pp(\bar p) \to t l \nu X$,      & \mcfm (\applgrid) \\
              &           &$pp(\bar p) \to tX$,             &  \\
              &           &$pp(\bar p) \to tWX$             &  \\ [0.5ex]
\cline{2-4}  \\ [-2.0ex]
             &jets &$pp(\bar p) \to \mathrm{jets} X$ & \nlojetpp (\applgrid), \\
                &  & & \nlojetpp (\fastnlo) \\ [0.5ex]
\hline  \\ [-2.5ex] 
LHC& DY heavy  &$pp \to VhX$ & \mcfm (\applgrid) \\  [0.5ex]
& quarks  & & \\  [0.5ex]
\hline
\end{tabular}
\caption{The list of experimental data and theory calculations implemented in the \fitter package. 
The references for the individual calculations and schemes are given in the text.
}
\label{tab:proc}
\end{table}
\normalsize
\paragraph{Theory:}  
 The PDFs are parametrised at a starting scale, $Q_0^2$,  
using a functional form and a set of free parameters $\vec{p}$. 
These PDFs are evolved to the scale of the measurements $Q^2$, $Q^2>Q_0^2$.
By default, the evolution uses the DGLAP formalism \cite{Gribov:1972ri, Gribov:1972rt, Lipatov:1974qm,
Dokshitzer:1977sg, Altarelli:1977zs} as implemented in \qcdnum~\cite{qcdnum}. Alternatively, 
the CCFM evolution \cite{\CCFM} as implemented in \texttt{uPDFevolv}~\cite{tmdlref2} can be chosen.
The prediction of the cross section for a particular process is obtained, assuming factorisation, by the convolution of the evolved 
PDFs with the corresponding parton scattering cross section.
Available theory calculations for each process are listed in Tab.~\ref{tab:proc}.
Predictions using dipole models~\cite{Golec-Biernat:1998js,Iancu:2003ge,Bartels:2002cj} can also be obtained.
\paragraph{QCD Analysis:} \rm  
The PDFs are determined in a least squares fit: a $\chi^2$ function, which compares the input data and theory predictions, 
is minimised with the \minuit \cite{minuit} program.
%
In \fitter various choices are available for the treatment of experimental uncertainties in the $\chi^2$ definition. 
Correlated experimental uncertainties can be accounted for using 
a nuisance parameter method  
or a covariance matrix method as described in Sec.~\ref{sec:chi2representation}.  Different statistical assumptions for the distributions of the systematic uncertainties, e.g. Gaussian or LogNormal~\cite{hera-lhc:report2009}, can also be studied (see Sec.~\ref{sec:experimentalerrors}).
\paragraph{Results:}
The resulting PDFs are provided in a format ready to be used by the \lhapdf 
library~\cite{lhapdf,lhapdfweb} or by \tmdlib \cite{Hautmann:2014kza}.
\fitter drawing tools can be used to display the PDFs with their uncertainties at a chosen scale.  
As an example, the first set of PDFs extracted using \fitter from HERA I data, HERAPDF1.0 \cite{h1zeus:2009wt}, 
is shown in Fig.~\ref{fig:hera1} (taken from Ref. \cite{h1zeus:2009wt}).
Note that following conventions, the PDFs are displayed as parton momentum distributions $xf(x,\mu_F^2)$.

\section{Theoretical formalism using DGLAP evolution}
\label{sec:theory}

%
In this section the theoretical formalism based on DGLAP \cite{Gribov:1972ri,Gribov:1972rt,Lipatov:1974qm,Dokshitzer:1977sg,Altarelli:1977zs} equations
is described. 

A direct consequence of factorisation (Eq. \ref{eq:fact}) is that the scale dependence or ``evolution'' of the PDFs can be predicted 
by the renormalisation group equations. 
By requiring physical observables to be independent of 
$\muf$, a representation 
of the parton evolution in terms of the DGLAP equations is obtained:

\begin{eqnarray}
\frac{d~f_a(x,\muf^2)}{d \log\muf^2} = \sum_{b=q,\bar{q},g}\int_{x}^{1}\frac{d z}{z} 
P_{ab}\left(\frac{x}{z};\muf^2\right) f_b(z,\muf^2)\,,
\label{eq:dglap}
\end{eqnarray}
where the functions $P_{ab}$ are the evolution kernels or splitting functions, which represent the probability 
of finding parton $a$ in parton $b$. They can be calculated as a  perturbative expansion in $\alpha_s$. 
Once PDFs are determined at the initial
scale $\mu_F^2 = Q_0^2$, their evolution to any other scale $Q^2 > Q_0^2$ is entirely determined by the DGLAP equations.
The PDFs are then used to calculate cross sections for various different processes.
Alternative approaches to DGLAP evolution equations, valid in different kinematic regimes, 
are also implemented in \fitter and will be discussed in Sec.~\ref{sec:alternative}.


\subsection{Deep Inelastic Scattering and Proton Structure}
\label{dissection}

The for\-ma\-lism that relates the DIS measurements to pQCD and the PDFs has been described
in detail in many extensive reviews (see e.g. Ref. \cite{disbook}) and it is only briefly summarised here.
DIS is the process where a lepton scatters off the partons in the proton
by the virtual exchange of a neutral ($\gamma/Z$) or charged ($W^{\pm}$) vector boson and, as a result, a scattered lepton and a 
hadronic final state are produced.
The common DIS kinematic variables are the scale of the process $Q^2$, which is the absolute squared four-momentum of 
the exchanged boson, Bjorken $x$, 
which can be related in the parton model to 
the momentum fraction that is carried by the struck quark, 
and the inelasticity $y$. These are related by $y=Q^2/sx$, where $s$ is the squared centre-of-mass energy.
\\
The NC cross section can be expressed in terms of generalised structure functions:
\begin{eqnarray}
    \frac{d^2\sigma_{NC}^{e^{\pm} p}}{dxdQ^2}&=&\frac{2\pi\alpha^2 Y_{+}}{xQ^4} \sigma_{r,NC}^{e^{\pm} p},\\ 
    \sigma_{r,NC}^{e^{\pm} p}&= &  \tilde F_2^{\pm} \mp \frac{Y_{-}}{Y_{+}}x \tilde F_3^{\pm} - \frac{y^2}{Y_{+}} \tilde F_L^{\pm},
\end{eqnarray}
where  $Y_{\pm} = 1 \pm (1-y)^2$ and $\alpha$ is the electromagnetic coupling constant.
The generalised structure functions $\tilde F_{2,3}$ 
can be written as linear combinations of the proton structure functions $F^{\gamma}_2, F^{\gamma Z}_{2,3}$ 
and $F^Z_{2,3}$, which are associated with pure photon exchange terms, photon-$Z$ interference
terms and pure $Z$ exchange terms, respectively. 
The structure function $\tilde F_2$ is the dominant contribution to the cross section, 
$x \tilde F_3$ becomes important at high $Q^2$ and $\tilde F_L$ is sizable 
only at high $y$. 
In the framework of pQCD, the structure functions are directly related to the 
PDFs: at LO $F_2$ is the weighted momentum sum of quark and anti-quark distributions, 
$F_2~\approx~x \sum e^2_q (q+ \overline q)$ (where $e_q$ is the quark electric charge),
$xF_3$ is related to their difference, 
$xF_3~\approx~x \sum 2e_q a_q (q- \overline q)$ ($a_q$ is the axial-vector quark coupling),
and $F_L$ vanishes. At higher orders, terms related to the gluon distribution
appear, in particular $F_L$ is strongly related to the low-$x$ 
gluon.
\\
The inclusive CC $ep$ cross section, analogous to the NC $ep$ case, can be expressed in terms of another set 
of structure functions, $\tilde W$: 
\begin{eqnarray}
\centering
   \frac{d^2\sigma_{CC}^{e^{\pm} p}}{dxdQ^2}&=&\frac{1\pm P}{2}\frac{G^2_F}{2\pi x} \ \frac{m^2_W}{m^2_W+Q^2} \ \sigma_{r,CC}^{e^{\pm} p}\\
   \sigma_{r,CC}^{e^{\pm} p}&= &  Y_{+} \tilde W_2^{\pm} \mp Y_{-}x \tilde W_3^{\pm} - y^2 \tilde W_L^{\pm},
\end{eqnarray}
where $P$ represents the lepton beam polarisation.
At LO in $\alpha_s$, the CC $e^+p$ and $e^-p$ cross sections are sensitive to 
different combinations of the quark flavour densities:
 \begin{eqnarray}
     \sigma_{r,CC}^{e^{+} p} &\approx& 
       x [\overline u + \overline c] + (1-y)^2 x [d+s], \\
     \sigma_{r,CC}^{e^{-} p} &\approx& 
       x[u+c] + (1-y)^2 x[\overline d + \overline s].
 \end{eqnarray}

Beyond LO, the QCD predictions for the DIS structure functions are obtained by convoluting 
the PDFs with appropriate hard-process scattering matrix elements, which are referred to as coefficient functions.

%

The DIS measurements span a large range of $Q^2$ from a few $\GeV^2$ to about $10^5$ $\GeV^2$, crossing heavy quark mass thresholds, thus the treatment of heavy quark (charm and beauty) 
production and the chosen values of their masses become important. 
There are different schemes for the treatment of heavy quark production. 
Several variants of these schemes are implemented in \fitter and they are briefly discussed below.

\paragraph{Zero-Mass Variable Flavour Number (ZM-VFN)\rm:\\}
In this scheme~\cite{ZMVFNpub}, the
heavy quarks appear as partons in the proton at $Q^2$ values above $\sim m_h^2$ (heavy quark mass)
and they
are then treated as massless in both the initial 
and final states of the hard scattering process. The lowest order process is the
scattering of the lepton off the heavy quark via electroweak boson exchange.
This scheme is expected to be reliable only in the region where $Q^2\gg m_h^2$, and it is inaccurate for lower 
$Q^2$ values since it misses corrections of order $m_h^2/Q^2$, while the other schemes mentioned 
below are accurate up to order $\Lambda_{\rm QCD}^2/Q^2$ albeit with different perturbative orderings.
In \fitter this scheme is available for the DIS structure function calculation 
via the interface to the \qcdnum \cite{qcdnum} package, thus it benefits 
from the fast \qcdnum convolution engine.

\paragraph{Fixed Flavour Number (FFN)\rm:\\} 
In this rigorous quantum field theory scheme \cite{Laenen:1992, Laenen:1993, Riem:1995}, 
only the gluon and the light quarks are considered
as partons within the proton and massive 
quarks are produced perturbatively in the final state.
The lowest order process is
the heavy quark-antiquark pair production via boson-gluon fusion.
In \texttt{HERA}\texttt{Fitter} this scheme can be accessed via the 
\qcdnum implementation or through the interface to the open-source code \texttt{OPENQCDRAD}~\cite{openqcdrad:page} as implemented by the ABM group.
This scheme is reliable only for $Q\sim m_h^2$, since it does not resum logarithms of the 
form $\ln(Q^2/m_h^2)$ which become important for $Q^2\gg m_h^2$.
In \qcdnum, the calculation of the heavy quark contributions to DIS structure functions
are available at Next-to-Leading Order (NLO) and only electromagnetic exchange contributions are taken into account. 
In the \texttt{OPEN}\texttt{QCDRAD} implementation the heavy quark contributions to CC structure functions are also available 
and, for the NC case, the QCD corrections to the coefficient functions in Next-to-Next-to Leading Order (NNLO)
are provided in the best currently known approximation~\cite{SMoch:npb864,Bierenbaum:2009mv}.
The  \texttt{OPENQCDRAD} implementation uses in addition the running heavy quark mass in the $\overline{\text{MS}}$ scheme~\cite{Alekhin:runm}.
It is sometimes argued that this $\overline{\text{MS}}$ scheme reduces the sensitivity of the DIS cross sections to higher order 
corrections. 
It is also known to have smaller non-perturbative corrections than the pole mass scheme \cite{Beneke:1998ui}.

\paragraph{General-Mass Variable Flavour Number (GM-VFN)\rm:\\}
In this scheme (see \cite{Thorne:2008xf} for a comprehensive review), heavy quark production is treated for
$Q^2 \sim m_h^2$ in the FFN scheme and for $Q^2 \gg m_h^2$
in the massless scheme with a suitable interpolation in between. 
The details of this interpolation differ between implementations.
The groups that use GM-VFN schemes in PDFs are MSTW, CT (CTEQ), NNPDF, and HERAPDF.
\fitter implements different variants of the GM-VFN scheme.
\begin{itemize}
\item \it {GM-VFN Thorne-Roberts scheme:} \rm
%
%
The Thorne-Roberts (TR) scheme~\cite{Thorne:1997ga} was designed to provide a smooth transition 
from the massive FFN scheme at low scales $Q^2 \sim m_h^2$ to the massless ZM-VFNS scheme at high scales $Q^2 \gg m_h^2$. 
Because the original version was technically difficult to implement beyond NLO, it was updated 
to the TR$^\prime$ scheme~\cite{Thorne:2006qt}.
There are two variants of the TR$^\prime$ schemes: TR$^\prime$ standard (as used in MSTW PDF sets~\cite{Thorne:2006qt,MSTWpdf}) 
and TR$^\prime$ optimal~\cite{Thorne:6180}, with a smoother transition across the heavy quark threshold region. 
Both TR$^\prime$ variants are accessible within the \fitter package at LO, NLO and NNLO.
At NNLO, an approximation is needed for the massive $\mathcal{O}(\alpha_s^3)$ NC coefficient 
functions relevant for $Q^2\sim m_h^2$, as for the FFN scheme.
\vspace{0.1cm}
\item \it {GM-VFN ACOT scheme:} \rm
The Aivazis-Collins-Olness-Tung (ACOT) scheme belongs to the group of VFN factorisation 
schemes that use the renormalisation method of Collins-Wilczek-Zee (CWZ) \cite{CWZ}.
This scheme unifies the low scale $Q^2 \sim m_h^2$ and high scale $Q^2 > m_h^2$ regions
in a coherent framework across the full energy range.
Within the ACOT package, the following variants of the ACOT $\overline{\text{MS}}$ scheme are available at LO and NLO:
ACOT-Full \cite{Aivazis:1993pi}, S-ACOT-$\chi$ \cite{Kramer:2000hn,Kretzer:2003it} and ACOT-ZM \cite{Aivazis:1993pi}.
For the longitudinal structure function higher order calculations are also available. 
A comparison of PDFs extracted from QCD fits to the HERA data 
with the TR$^\prime$ and ACOT-Full schemes is illustrated in Fig.~\ref{fig:acotrt} (taken from \cite{h1zeus:2009wt}).

\begin{figure}[!ht]
\centering
\includegraphics[width=8cm]{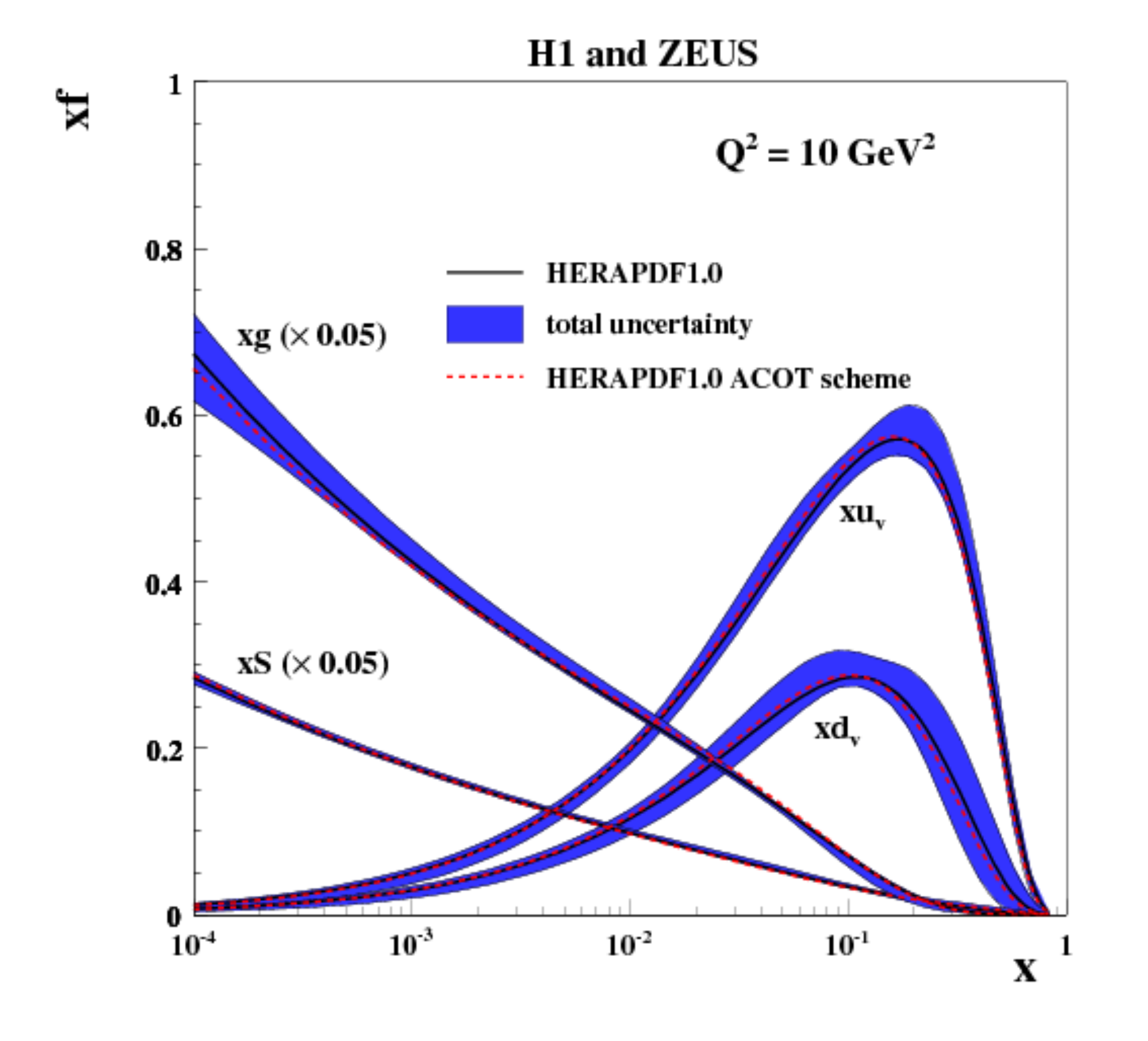}
   \caption{Distributions of valence ($xu_v$, $xd_v$), sea ($xS$) and the gluon ($xg$) PDFs in HERAPDF1.0~\cite{h1zeus:2009wt}
       with their total uncertainties at the scale of $Q^2 = 10\ \GeV^2$ obtained 
       using the TR$^\prime$ scheme and compared to the PDFs obtained with 
       the ACOT-Full scheme using the $k$-factor technique (red).
       The gluon and the sea distributions are scaled down by a factor of 20.}
 \label{fig:acotrt}
\end{figure}

\end{itemize}

\subsection{Electroweak Corrections to DIS}
Calculations of higher-order electroweak corrections to DIS at 
HERA are available in \fitter in the on-shell scheme. In this scheme, the
masses of the gauge bosons $m_W$ and 
$m_Z$ are treated as basic parameters together with the top, 
Higgs and fermion masses.
These electroweak corrections 
are based on the \texttt{EPRC} package~\cite{SpiesbergerPrivComm}.
The code calculates the running of the electromagnetic coupling $\alpha$ using the most recent parametrisation
of the hadronic contribution
\cite{Jegerlehner} as well as 
an older version from Burkhard \cite{Burkhard}.

\subsection{Diffractive PDFs}

\newcommand{\asotp}{\ensuremath{\frac{\alpha_{\rm s}}{2\pi}}}
\newcommand{\Sgl}[1]{\ensuremath{\tilde f_{#1+}}}
\newcommand{\Pom}{{I\!P}}
\newcommand{\Reg}{{I\!R}}
\newcommand{\xpom}{$x_{I\!P}$}
\newcommand{\xP}{x_\Pom}

About 10\% of deep inelastic interactions at HERA are diffractive, such that the interacting proton stays intact ($ep\to eXp$). 
The outgoing proton is separated from the rest of the final hadronic system, $X$, by a large rapidity gap.  
Such events are a subset of DIS where the hadronic state $X$ comes from the interaction of the
virtual photon with a colour-neutral cluster stripped off the proton \cite{Hebecker:1999vf}.
The process can be described analogously to the inclusive DIS, by means of the diffractive parton distributions 
(DPDFs) \cite{Collins:1997sr}.
The parametrization of the colour-neutral exchange in terms of factorisable `hard' Pomeron and a secondary 
Reggeon \cite{Ingelman:1984ns}, both having a hadron-like partonic structure, 
has proved remarkably successful in the description of most of the diffractive data.
It has also provided a practical method to determine DPDFs from fits to the diffractive cross sections.

%

In addition to the usual DIS variables $x$, $Q^2$, extra kinematic variables are needed to describe the diffractive process. 
These are the squared four-momentum transfer of the exchanged Pomeron or Reggeon, $t$, and 
the mass $m_X$ of the diffractively produced final state. 
In practice, the variable $m_X$ 
is often replaced by the dimensionless quantity $\beta=\frac{Q^2}{m_X^2+Q^2-t}$.
In models based on a factorisable Pomeron, $\beta$ may be viewed at LO as the fraction of the
Pomeron longitudinal momentum, $x_{\Pom}$, which is carried by the struck parton, $x=\beta x_{\Pom}$,
where $P$ denotes the momentum of the proton.
\\
For the inclusive case, the diffractive cross section reads as:
\begin{equation}
\begin{array}{lcl}
    \frac{d^4\sigma}{d\beta\,dQ^2dx_{\Pom}\,dt}
=
  \frac{2\pi\alpha^2}{\beta Q^4}\,
    \left( 1 +  (1-y)^2 \right) \ensuremath{\overline\sigma}^{D(4)}(\beta,Q^2,x_{\Pom},t)
\label{Dxs}
\end{array}
\end{equation}
with the ``reduced cross section'': 
\begin{equation}
\begin{array}{lcl}
\overline\sigma^{D(4)}
 = F_2^{D(4)} - \frac{y^2}{1 +  (1-y)^2}\, F_L^{D(4)}.
\label{eq:sigred}
\end{array}
\end{equation}

The diffractive structure functions can be expressed as convolutions of 
calculable coefficient functions with the diffractive quark and gluon distribution functions,
 which in general depend on \xpom, $Q^2$, $\beta$ and $t$.

The DPDFs \cite{Aktas:2006hy, zeus:diff2009} in \fitter are implemented as a sum 
of two factorised contributions:
\begin{equation}
 \Phi_\Pom(\xP,t)\, f^{\Pom}_{a}(\beta,Q^2)
  + 
 \Phi_\Reg(\xP,t)\, f^{\Reg}_{a}(\beta,Q^2)
 \,,
\end{equation} 
where $\Phi(\xP,t)$ are the Reggeon and Pomeron fluxes.
The Reggeon PDFs, $f^{\Reg}_{a}$ are fixed as those of the pion, while the Pomeron PDFs,
$f^{\Pom}_{a}$, can be obtained from a fit to the data.

%

\subsection{Drell-Yan Processes in $pp$ or $p\bar p$ Collisions}
\label{dysection}

The Drell-Yan (DY) process
provides valuable information about PDFs.
In $pp$ and $p\bar p$ scattering, the $Z/\gamma^*$ and $W$ production 
probe bi-linear combinations of quarks. 
Complementary information on the different quark densities
can be obtained from the $W^{\pm}$ asymmetry ($d$, $u$ and their ratio),
the ratio of the $W$ and $Z$ cross sections (sensitive to the flavour 
composition of the quark sea, in particular to the $s$-quark distribution), 
and associated $W$ and $Z$ production with
heavy quarks (sensitive to $s$, $c$- and $b$-quark densities).
 Measurements at large boson transverse momentum $p_T\gtrsim m_{W,Z}$ are potentially sensitive to the gluon 
distribution~\cite{Malik:2013kba}.
%


At LO the DY NC cross section triple differential in invariant mass \(m\), boson rapidity \(y\) 
and lepton scattering angle \(\cos\theta\) in the parton centre-of-mass frame can be written as~\cite{Drell:1970wh,Yamada:1981mw}:
\begin{eqnarray}
 \frac{d^3\sigma}{dm{d}y d\cos\theta} &=&  
 \frac{\pi\alpha^2}{3ms}\sum\limits_{q}\hat{\sigma}^{q}(\cos\theta, m)  \nonumber \\
 &\times &\left[f_q(x_1,m^2)f_{\bar{q}}(x_2,m^2) 
 + (q\leftrightarrow\bar{q})\right],
\end{eqnarray}
where \(s\) is the squared centre-of-mass beam energy, the parton momentum fractions are given by \(x_{1,2} = \frac{m}{\sqrt{s}}\exp(\pm y)\), $f_q(x_1,m^2)$ 
are the PDFs at the scale of the invariant mass, and 
$\hat{\sigma}^{q}$ is the parton-parton hard scattering cross section. 
%
\\
\\
The corresponding triple differential CC cross section has the form:
\begin{eqnarray}
\frac{d^3\sigma}{dmdyd\cos\theta} &=&
 \frac{\pi\alpha^2}{48s\sin^4\theta_W}
 \frac{m^3(1-\cos\theta)^2}{(m^2-m_W^2)+\Gamma_W^2m_W^2}  \nonumber \\
 &\times& \sum_{q_1,q_2}V_{q_1q_2}^2f_{q_1}(x_1,m^2)f_{q_2}(x_2,m^2),
\end{eqnarray}
where \(V_{q_1q_2}\) is the Cabibbo-Kobayashi-Maskawa (CKM) quark mixing matrix and \(m_W\) and \(\Gamma_W\)
are the \(W\) boson mass and decay width, respectively.

The simple LO form of these expressions allows for the analytic calculations of integrated
cross sections.
In both NC and CC expressions the PDFs depend only on the boson rapidity \(y\) and
invariant mass \(m\), while
the integral in \(\cos\theta\) can be evaluated analytically
even for the case of realistic kinematic cuts.
%

Beyond LO, the calculations are often time-consuming and Monte Carlo generators are employed. 
Currently, the predictions for $W$ and $Z/\gamma^*$ production are available up
to NNLO and the predictions for $W$ and $Z$ production in association with heavy flavour quarks are available to NLO.

There are several possibilities to obtain the theoretical
predictions for DY production in \fitter. 
The NLO and NNLO calculations can be implemented using $k$-factor or \emph{fast grid} techniques (see Sec.~\ref{sec:techniques}
for details), which are interfaced to programs such as
\texttt{MCFM}~\cite{Campbell:1999ah,Campbell:2000je,Campbell:2010ff}, 
available for NLO calculations, or 
\tt FEWZ\rm~\cite{FEWZ} and \tt DYNNLO\rm \cite{DYNNLO} for NLO and NNLO, with electroweak corrections estimated using \tt MCSANC\rm~\cite{Bardin:2012jk, Bondarenko:2013nu}.




\subsection{Jet Production in $ep$ and $pp$ or $p \bar p$ Collisions}
\label{jetsection}

The cross section for production of high $p_T$ hadronic jets
is sensitive to the high-$x$ gluon 
PDF (see e.g. Ref.~\cite{MSTWpdf}). 
Therefore this process can be used to improve the determination of the gluon PDF,
which is particularly important for Higgs production and searches for new physics.
Jet production cross sections are currently known only to NLO.
Calculations for higher-order contributions to jet production in $pp$ collisions
are in progress~\cite{nigel:2013,nigel:2010,Currie:2013dwa}. 
Within \fitter, the \nlojetpp program \cite{Nagy:1998bb,Nagy:2001fj} may be used for 
calculations of jet production.
Similarly to the DY case, the calculation 
is very demanding in terms of computing power. 
Therefore \emph{fast grid} techniques are used  
to facilitate the QCD analyses including jet cross section measurements
in $ep$, $pp$ and $p\bar{p}$ collisions.
For details see Sec.~\ref{sec:techniques}.


\subsection{Top-quark Production in $pp$ or $p \bar p$ Collisions}

At the LHC, top-quark pairs ($t \bar t$) are produced dominantly via $gg$ fusion.
Thus, LHC measurements of the $t \bar t$ cross section provide additional 
constraints on the gluon distribution at medium to high values of $x$, 
on $\as$ and on the top-quark mass, $m_t$ \cite{cms:top}. 
Precise predictions for the total inclusive $t \bar t$ cross section are available 
up to NNLO~\cite{Czakon:2013goa} and they can be computed within \fitter via an interface 
to the program \texttt{HATHOR}~\cite{Aliev:2010zk}. 

Fixed-order QCD predictions for the differential $t \bar t$ cross section at NLO can be obtained by using
the program \texttt{MCFM}~\cite{Campbell:2010ff,Campbell:2009ss,Campbell:2005bb,Campbell:2004ch,Campbell:2012uf} 
interfaced to \fitter with \emph{fast grid} techniques.

Single top quarks are produced by exchanging electroweak bosons and the measurement of their production cross section can be used, for example, 
to probe the ratio of the $u$ and $d$ distributions in the proton 
as well as the $b$-quark PDF. Predictions 
for single-top production are available at the NLO accuracy by using \texttt{MCFM}.

Approximate predictions up to NNLO in QCD for the differential $t\bar{t}$ cross section in one-particle 
inclusive kinematics are available in \fitter through an interface to the program \difftop \cite{Guzzi:2014wia,difftop-web}.
It uses methods of QCD threshold resummation beyond the leading logarithmic approximation.
This allows the users to estimate the impact of the recent $t\bar{t}$ differential cross section measurements on the uncertainty 
of the gluon density within a QCD PDF fit at NNLO.
A fast evaluation of the \difftop differential cross sections is possible via an interface to \emph{fast grid}
computations \cite{dis2014Fast}. 

%

\section{Computational Techniques}
\label{sec:techniques}

Precise measurements
require accurate theoretical predictions in
order to maximise their impact in PDF fits.  Perturbative
calculations become more complex and time-consuming at higher  
orders due to the increasing number of relevant Feynman diagrams.
The direct inclusion of computationally
demanding higher-order calculations into iterative fits is thus
not possible currently. However, a full repetition of the
perturbative calculation for small changes in input parameters is
not necessary at each step of the iteration.
Two methods have been developed which take advantage of this
to solve the problem: the $k$-factor technique and the
\emph{fast grid} technique. Both are available in \fitter.

\subsection{$k$-factor Technique}
  The $k$-factors are defined as the ratio of the prediction of a
  higher-order (slow) pQCD calculation to a lower-order (fast)
  calculation using the same PDF. Because the $k$-factors depend on the phase space
  probed by the measurement, they have to be stored  
including their dependence on
  the relevant kinematic variables. Before the start of  
  a fitting procedure, a table of $k$-factors is computed once
  for a fixed PDF with the time consuming higher-order code. In
  subsequent iteration steps the theory prediction is derived from the
  fast lower-order calculation by multiplying by the pre-tabulated
  $k$-factors.

  This procedure, however, neglects the fact that the $k$-factors are 
  PDF dependent, and 
  as a consequence, they have to be re-evaluated
  for the newly determined PDF at the end of the fit for a consistency
  check. The fit must be repeated until input and output
  $k$-factors have converged. In summary, this technique avoids
  iteration of the higher-order calculation at each step, but still
  requires typically a few re-evaluations.

In \fitter, the $k$-factor technique can also be used for the fast 
computation of the time-consuming GM-VFN schemes for heavy quarks in DIS.
``FAST'' heavy-flavour schemes are implemented
with $k$-factors defined as the ratio of
calculations at the same perturbative order but for massive vs.\
massless quarks, e.g.\ NLO (massive)/NLO (massless).
These $k$-factors are calculated only for the
starting PDF and hence, the ``FAST'' heavy flavour schemes should
only be used for quick checks. Full heavy flavour schemes
should be used by default. However, for the ACOT scheme,
due to exceptionally long computation times, the $k$-factors are used in 
the default setup of \fitter. 

\subsection{\emph{Fast Grid} Techniques}

  \emph{Fast grid} techniques exploit the fact that iterative PDF fitting
  procedures do not impose completely arbitrary changes to the types
  and shapes of the parameterised functions that represent each PDF\@.
  Instead, it can be assumed that a generic PDF can be approximated by
  a set of interpolating functions with a sufficient number of
 judiciously chosen support points. The 
  accuracy of this approximation is checked and optimised 
 such that the approximation bias is negligibly
  small compared to the experimental and theoretical accuracy. 
   This method can be used to perform
  the time consuming higher-order calculations (Eq.~\ref{eq:fact})
  only once for the set of interpolating functions. 
  Further iterations of the calculation for
  a particular PDF set are fast, involving only sums over
  the set of interpolators multiplied by factors depending on the
  PDF\@. This approach can be used to calculate the cross sections 
  of processes involving one or two hadrons in the initial state and to
  assess their renormalisation and factorisation scale variation.

  This technique serves to facilitate the inclusion of time consuming
  NLO jet cross section predictions into PDF fits and has been implemented
  in the two projects, \fastnlo \cite{Adloff:2000tq,Kluge:2006xs} 
  and \applgrid \cite{Carli:2005ji,Carli:2010rw}.
  The packages differ in their interpolation
  and optimisation strategies, but both of them construct tables with
  grids for each bin of an observable in two steps: in the first step,
  the accessible phase space in the parton momentum fractions $x$ and
  the renormalisation and factorisation scales \mur and \muf is
  explored in order to optimise the table size. In the second step
  the grid is filled for the
  requested observables. Higher-order cross sections can then be
  obtained very efficiently from the pre-produced grids while varying
  externally provided PDF sets, \mur and \muf, or $\alpha_s(\mu_R)$. 
  This approach can in principle be extended to arbitrary
  processes. This requires an interface between the
  higher-order theory programs and the fast interpolation frameworks.
  For the \fitter implementations of the two packages, the evaluation of $\alpha_s$
  is done consistently with the PDF evolution code.
  A brief description of each package is given below:

\begin{itemize}
  \item The \fastnlo project~\cite{Kluge:2006xs} has been interfaced
    to the \nlojetpp program~\cite{Nagy:1998bb} for the calculation of
    jet production in DIS~\cite{Nagy:2001xb} as well as 2- and 3-jet
    production in hadron-hadron collisions at
    NLO~\cite{Nagy:2003tz,Nagy:2001fj}.  Threshold corrections at 2-loop
    order, which approximate NNLO for the inclusive jet cross
    section for $pp$ and $p\bar p$, have also been included into the framework \cite{Wobisch:2011ij} 
    following Ref.~\cite{Kidonakis:2000gi}.

    The latest version of the \fastnlo convolution program \cite{Britzger:2012bs} allows for the
    creation of tables in which renormalisation and factorisation scales
    can be varied as a function of two pre-defined observables, e.g.\ jet
    transverse momentum \pperp and $Q$ for DIS\@. 
    Recently, the differential calculation of top-pair production in hadron collisions 
    at approximate NNLO \cite{Guzzi:2014wia} has been interfaced to \fastnlo \cite{dis2014Fast}.
    The \fastnlo code is available online \cite{fastNLO:HepForge}.
    Jet cross section grids computed for the kinematics of various experiments
    can be downloaded from this site.

    The \fastnlo libraries and tables with theory predictions for comparison to
    particular cross section measurements are included in the \fitter package. 
    The interface to the \fastnlo tables from within \fitter was used in a recent
    CMS analysis, where the impact on extraction of the PDFs from the inclusive 
    jet cross section is investigated \cite{Khachatryan:2014waa}.
\\
\item In the \applgrid package~\cite{Carli:2010rw,APPLGRID:HepForge},
    in addition to jet cross sections for
    $pp(p\bar p)$ and DIS processes, calculations 
    of DY production and other processes are also implemented using an interface to the
    standard \mcfm parton level generator~\cite{Campbell:1999ah,Campbell:2000je,Campbell:2010ff}.
    Variation of the renormalisation and factorisation scales is possible a posteriori,
    when calculating theory predictions with the \applgrid  tables, and
    independent variation of $\alpha_S$ is also allowed.
    For predictions beyond NLO, the $k$-factors technique can also be applied
    within the \applgrid framework.

    As an example, the \fitter interface to \applgrid was used by the ATLAS~\cite{atlas:strange}
    and CMS~\cite{cms:strange} collaborations to extract the strange quark distribution of the proton.
    The ATLAS strange PDF extracted employing these techniques is displayed in
    Fig.~\ref{fig:atlas} together with a comparison to the global PDF
    sets CT10~\cite{CT10pdf} and NNPDF2.1 \cite{NNPDFpdf} (taken from \cite{atlas:strange}).

\begin{figure}[!ht]
  \centering
  \includegraphics[width=8cm]{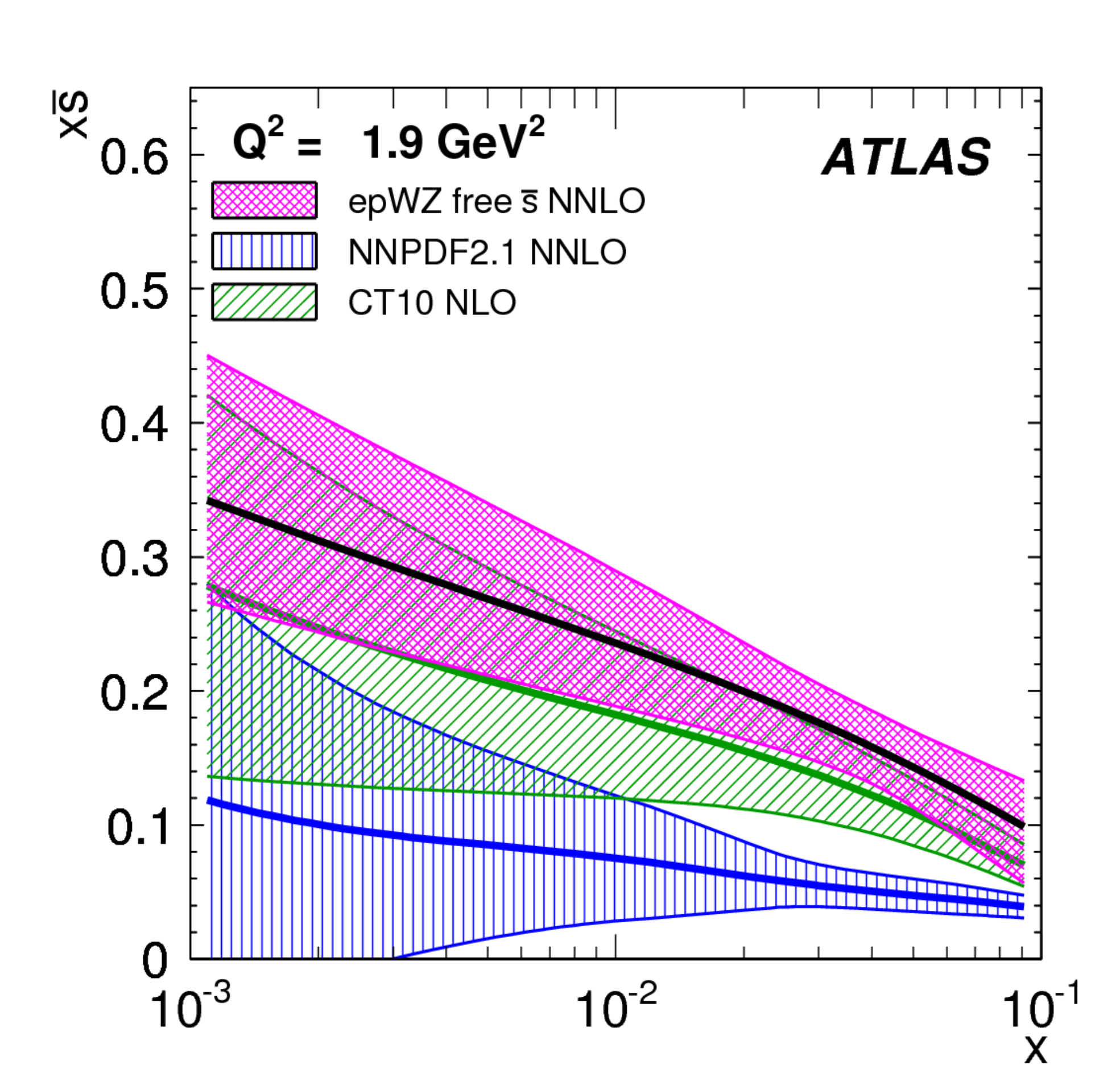}
  \caption{The strange antiquark distribution versus $x$ for the ATLAS
    epWZ free $\bar s$ NNLO fit \cite{atlas:strange} (magenta band) compared to predictions
    from NNPDF2.1 (blue hatched) and CT10 (green hatched) 
    at $Q^2 = 1.9\ \GeV^2$. The ATLAS fit was performed using a $k$-factor approach 
    for NNLO corrections.}
  \label{fig:atlas}
\end{figure}

\end{itemize}

\section{Fit Methodology}
\label{sec:method}


When performing a QCD analysis to determine PDFs there are various assumptions and choices to be made concerning, for example, the functional form of the input parametrisation, the treatment of heavy quarks and their mass values, alternative theoretical calculations, alternative representations of the fit $\chi^2$ and for different ways of treating correlated systematic uncertainties.
 It is useful to discriminate or quantify the effect of a chosen ansatz within a common framework and 
\fitter is optimally designed for such tests.
The methodology employed by \fitter  relies on a flexible and modular
framework that allows independent integration of state-of-the-art techniques, either related to the inclusion of a new theoretical calculation, or of new approaches to treat data and their uncertainties. 

In this section we describe the available options for the fit methodology in \fitter.
%
In addition, as an alternative approach to a complete QCD fit, the Bayesian reweighting
method, which is also available in \fitter, is described.
\subsection{Functional Forms for PDF Parametrisation}

\begin{figure}[!ht]
   \centering
   \includegraphics[width=8cm]{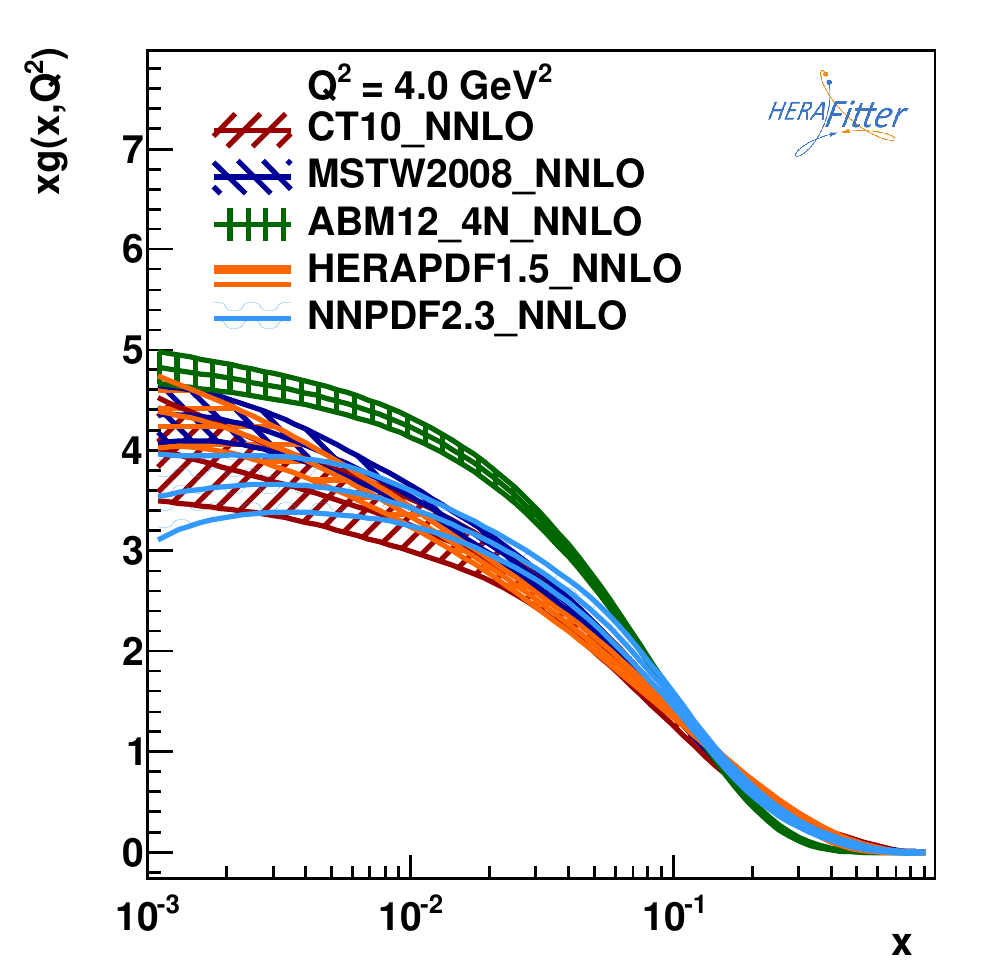}
   \caption{The gluon PDF as extracted by various groups at the scale of $Q^2=4 \ \GeV^2$, plotted using the drawing tools from \fitter.} 
 \label{fig:pdfs}
\end{figure}

Careful consideration must be taken when assigning the PDF freedom via functional forms. The PDFs can be parametrised using several predefined functional forms and 
flavour decompositions, as described briefly below. The choice of functional form can lead to a different shape for the PDF distributions, and consequently the size 
of the PDF uncertainties can depend on the flexibility of the parametric choice.
\paragraph{Standard Polynomials:} 
The standard polynomial form is the most commonly used. A polynomial functional form is used to parametrise the $x$-dependence of the PDFs, where index $j$ denotes each parametrised PDF flavour:
\begin{equation}
\centering
 xf_j(x) = A_j x^{B_j} (1-x)^{C_j} P_j(x).
\label{eqn:pdf_std}
\end{equation}
The parametrised PDFs are the valence distributions
$xu_v$ and $xd_v$, the gluon distribution $xg$, and the light sea quark distributions,
$x\bar{u}$, $x\bar{d}$,
$x\bar{s}$, 
at the starting scale, which is 
chosen below the charm mass threshold. 
The form of polynomials $P_j(x)$ can be varied.
The form $(1 + \epsilon_j \sqrt{x} + D_j x + E_j x^2)$
is used for the HERAPDF \cite{h1zeus:2009wt} 
with additional constraints relating to the flavour decomposition of the 
light sea. This parametrisation is termed HERAPDF-style. The polynomial can also
be parametrised in the CTEQ-style, where $P_j(x)$ takes the form $e^{a_3x} (1 + e^{a_4} x + e^{a_5} x^2)$ and,
in contrast to  the HERAPDF-style, this is positive by construction.
QCD number and momentum sum rules are used to determine the normalisations $A$ for the valence and gluon
distributions, and the sum-rule integrals are solved analytically.
\paragraph{Bi-Log-Normal Distributions:} 
This parametrisation is motivated by multi-particle statistics
and has the following functional form:
\begin{equation}
 xf_j(x)=a_j x^{p_j-b_j\log(x)}(1-x)^{q_j-d_j \log(1-x)}.
\label{eqn:bilog}
\end{equation}
This function can be regarded as a generalisation of the standard polynomial form described above,
however, numerical integration of Eq.~\ref{eqn:bilog} is required in order to impose the QCD sum rules.
\paragraph{Chebyshev Polynomials:} 
A flexible parametrisation  based on the Chebyshev polynomials can be employed for the gluon and sea distributions.
Polynomials with argument $\log(x)$ are considered for better modelling the low-$x$ asymptotic behaviour of those PDFs. 
The polynomials are multiplied
by a factor of $(1-x)$ to ensure that they vanish as $x\to 1$. The resulting parametric form reads
\begin{eqnarray}
x g(x) &=& A_g \left(1-x\right) \sum_{i=0}^{N_g-1} A_{g_i} T_i \left(-\frac{\textstyle 2\log x - \log x_{\rm min} } {\textstyle \log x_{\rm min} } \right)\,, \label{eq:glu} \\
x S(x) &=& \left(1-x\right) \sum_{i=0}^{N_S-1} A_{S_i} T_i \left(-\frac{\textstyle 2\log x - \log x_{\rm min} } {\textstyle \log x_{\rm min} } \right)\,, \label{eq:sea} 
\end{eqnarray}
where $T_i$ are first-type Chebyshev polynomials of order $i$.
The normalisation factor $A_g$ is derived from the momentum sum rule analytically.
Values of $N_{g,S}$ to $15$ are allowed, however the fit quality is already similar
to that of the standard-polynomial parametrisation from $N_{g,S} \ge 5$  and has a similar number of free parameters \cite{Chebyshev}.
\paragraph{External PDFs:} \rm 
 \fitter also provides the possibility to access external PDF sets, which can be used to compute 
theoretical predictions for the cross sections for all the processes available in \fitter. 
This is possible via an interface to \lhapdf~\cite{lhapdf,lhapdfweb} providing access to the 
global PDF sets.
\fitter also allows one to evolve PDFs from \lhapdf using \qcdnum.
Fig. \ref{fig:pdfs} illustrates a comparison of various gluon PDFs accessed from \lhapdf as produced with the drawing 
tools available in \fitter.
%
%
\subsection{Representation of \texorpdfstring{$\chi^2$}{χ²}}
\label{sec:chi2representation}

The PDF parameters are determined in \fitter by minimisation of a
$\chi^2$ function taking into account correlated and uncorrelated measurement uncertainties.
There are various forms of $\chi^2$, 
e.g. using a covariance matrix or providing nuisance parameters to encode the dependence of 
each correlated systematic uncertainty for each measured data point.
The options available in \fitter are the following:
\begin{description}
\item \it {Covariance Matrix Representation:} \rm
For a data point $\mu_i$ with a corresponding theory prediction $m_i$, 
the $\chi^2$ function 
can be expressed in the following form:
\begin{eqnarray}
\chi^2 (m)& = & \sum_{i,k}(m_i-\mu_i)C^{-1}_{ik}(m_k-\mu_k),
\end{eqnarray}
where the experimental uncertainties are given as a covariance matrix $C_{ik}$ for measurements in bins $i$ and $k$.
The covariance matrix $C_{ik}$ is given by a sum of statistical, uncorrelated and correlated systematic contributions: 
\begin{eqnarray}
C_{ik}& = & C^{stat}_{ik}+C^{uncor}_{ik}+C^{sys}_{ik}.
\end{eqnarray}
Using this representation one cannot distinguish the effect of each source of systematic uncertainty. 
\item \it{Nuisance Parameter Representation:} \rm
In this case, the $\chi^2$ is expressed as
\label{sec:nuisance_representation}
\begin{equation} 
    \chi^2\left(\boldsymbol{m},\boldsymbol{b}\right) =   
 \sum_i \frac{\left[ {\mu_i} - m_i \left( 1 - \sum_j \gamma^i_j b_j \right) \right]^2}
{ \textstyle \delta^2_{i,{\rm unc}}m_i^2 + \delta^2_{i,{\rm stat}}\, {\mu_i} m_i \left(1 - \sum_j \gamma^i_j b_j\right) }
  + \sum_j b^2_j,
\label{eq:aven}
\end{equation}
where, $\delta_{i,\rm stat}$ and $\delta_{i,\rm unc}$ are 
relative statistical and uncorrelated systematic uncertainties
of the measurement $i$.
Further, $\gamma^i_j$ quantifies the sensitivity of the
measurement to the correlated systematic source $j$. 
The function $\chi^2$ depends on
 the set of systematic nuisance parameters $b_j$.
This definition of the $\chi^2$ function assumes that
systematic uncertainties are proportional to the central prediction values
(multiplicative uncertainties, $m_i(1-\sum_j\gamma_j^ib_j)$), whereas the statistical uncertainties scale 
with the square root of the expected number of events. 
However, additive treatment of uncertainties is also possible in \fitter.

During the $\chi^2$ minimisation, the nuisance parameters $b_j$ and the PDFs are determined, such that the effect of different sources of systematic uncertainties can be distinguished.
\item  \it{Mixed Form Representation:} \rm
In some cases, the statistical and systematic uncertainties of experimental data are provided in different forms.    
For example, the correlated experimental systematic uncertainties are available as nuisance parameters,
but the bin-to-bin statistical correlations are given in the form of a covariance matrix.
\fitter\ offers the possibility to include such mixed forms of information. 
\end{description}
Any source of measured systematic uncertainty can be treated as additive or multiplicative, as described above. 
The statistical uncertainties can be included as additive or following the Poisson statistics. Minimisation
with respect to nuisance parameters is performed analytically, however, for more 
detailed studies of correlations individual nuisance parameters can be included into 
the \minuit minimisation.

\subsection{Treatment of the Experimental Uncertainties}
\label{sec:experimentalerrors}

Three distinct methods for propagating experimental uncertainties to PDFs are implemented in \fitter and reviewed here:
the Hessian, Offset, and Monte Carlo method.
\begin{description}
\item \it{Hessian (Eigenvector) method:} \rm
The PDF uncertainties reflecting the data experimental uncertainties are estimated by 
examining the shape of the $\chi^2$ function in the neighbourhood of the minimum \cite{Pumplin:2001ct}.
Following the approach of Ref. \cite{Pumplin:2001ct}, the Hessian matrix is defined by the second 
derivatives of $\chi^2$ on the fitted PDF parameters. The matrix is diagonalised and the 
Hessian eigenvectors are computed. 
Due to orthogonality these vectors correspond to independent sources of
uncertainty in the obtained PDFs.
\\
%

\item \it{Offset  method:} \rm
The Offset method \cite{Botje:2001fx} uses
%
the $\chi^2$ function for the central fit, but only
uncorrelated uncertainties are taken into account. 
The goodness of the fit can no longer be judged from the $\chi^2$ since correlated uncertainties are ignored. 
The correlated uncertainties are propagated into the PDF uncertainties by performing variants 
of the fit with the experimental data varied by $\pm 1 \sigma$ from the central value  
for each systematic source.
The resulting deviations of the PDF parameters from the ones obtained in the central 
fit are statistically independent, and they can be combined in quadrature to derive a total 
PDF systematic uncertainty.
%

The uncertainties estimated by the offset method are generally larger than 
those from the Hessian method.
\\

\item \it{Monte Carlo method:} \rm
The Monte Carlo (MC) technique \cite{Giele:1998gw, mcmethod2} can also be used to determine PDF uncertainties.
The uncertainties are estimated using pseudo-data replicas (typically $>100$) 
randomly generated from the measurement central values and their systematic and statistical uncertainties 
taking into account all point-to-point correlations.
The QCD fit is performed for each replica and the PDF central values and their 
experimental uncertainties are estimated from the distribution of the PDF parameters obtained in these fits, by taking 
the mean values and standard deviations over the replicas.
%

The MC method has been checked against the standard error estimation of the PDF uncertainties obtained by the Hessian method. 
A good agreement was found between the methods provided that Gaussian distributions of statistical and systematic uncertainties are assumed in the MC 
approach~\cite{hera-lhc:report2009}.
A comparison is illustrated in Fig.~\ref{fig:mchessian}. 
Similar findings were reported by the MSTW global analysis~\cite{Watt:2012tq}. 
\begin{figure}[!ht]
 \centering
  \includegraphics[trim=1cm 4cm 1cm 5cm, clip, width=6.3cm]{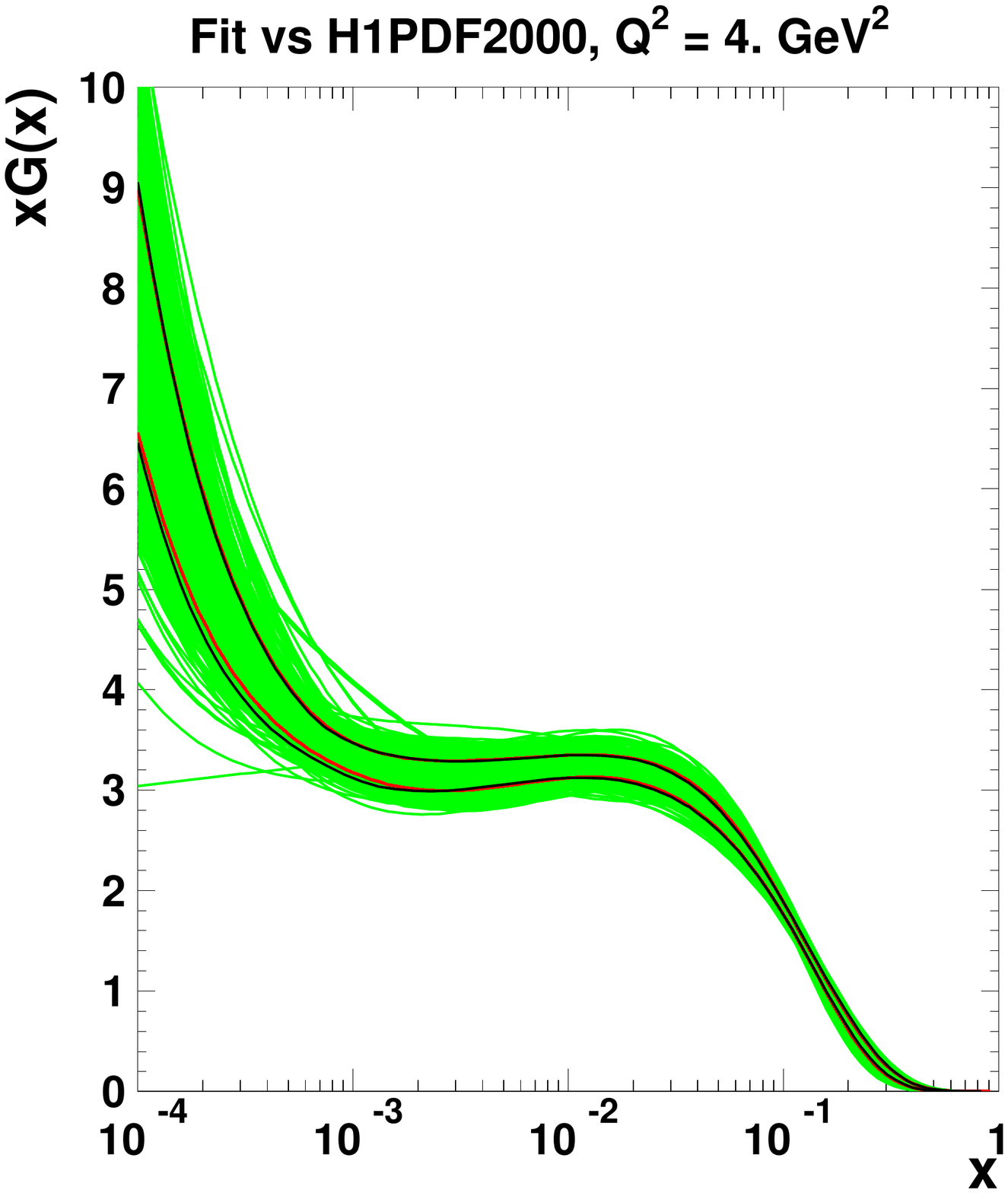}
  \caption{Comparison between the standard error calculations as employed by the Hessian approach (black lines) 
           and the MC approach (with more than 100 replicas) assuming Gaussian distribution for uncertainty 
           distributions, shown here for each replica 
          (green lines) together with the evaluated standard deviation (red lines) \cite{hera-lhc:report2009}.
          The black and red lines in the figure are superimposed because agreement of the methods is so good that it is hard to distinguish them.}
  \label{fig:mchessian}        
\end{figure}

Since the MC method requires large number of replicas, the eigenvector representation 
is a more convenient way to store the PDF uncertainties.
It is possible to transform MC to eigenvector
representation as shown by \cite{Gao:2013bia}. Tools to perform this transformation are provided with \fitter and were 
recently employed for the representation of correlated sets of PDFs at different perturbative orders \cite{hfcorrpaper}.
\end{description}
The nuisance parameter representation of $\chi^2$ in Eq.~\ref{eq:aven} is derived assuming 
symmetric experimental errors, however, the published systematic uncertainties are 
often asymmetric.
\fitter provides the possibility to use asymmetric systematic uncertainties.
The implementation relies on the assumption that 
asymmetric uncertainties can be described by a parabolic function.
The nuisance parameter in Eq.~\ref{eq:aven} is modified as follows
\begin{equation}
  \gamma^i_{j} \to \omega^i_{j}b_j + \gamma^i_{j},
\end{equation}
where the coefficients $\omega^i_{j}$, $\gamma^i_{j}$ are defined  
from the maximum and minimum shifts of the cross sections due to a  variation of the systematic uncertainty $j$, 
$S_{ij}^{\pm}$,
\begin{equation}
  \omega^i_{j}=\frac{1}{2}\left(S_{ij}^{+}+S_{ij}^{-}\right), \\
  \gamma^i_{j}=\frac{1}{2}\left(S_{ij}^{+}-S_{ij}^{-}\right). 
\end{equation}

\subsection{Treatment of the Theoretical Input}
\label{sec:theoryerr}

The results of a QCD fit depend not only on the input data but also on the 
input parameters used in the theoretical calculations. Nowadays, PDF groups 
address the impact of the choices of theoretical parameters
by providing
alternative PDFs with different choices of the mass of the charm quarks, $m_c$, 
mass of the bottom quarks, $m_b$, and the value of $\asmz$. 
Other important aspects are the choice of the functional form for the PDFs at the 
starting scale and the value of the starting scale itself. 
\fitter provides the
possibility of different user choices of all these inputs.

\subsection{Bayesian Reweighting Techniques}

As an alternative to performing a full QCD fit, \fitter allows the user to assess the impact of including new
data in an existing fit using the Bayesian Reweighting technique. The method
provides a fast estimate of the impact of new data on PDFs. 
Bayesian Reweighting was first proposed for PDF sets delivered in the form of MC replicas by~\cite{Giele:1998gw} 
and further developed by the NNPDF Collaboration~\cite{Ball:2011gg,Ball:2010gb}. 
More recently, a method to perform Bayesian Reweighting studies starting from PDF fits for which uncertainties
are provided in the eigenvector representation has been also developed~\cite{Watt:2012tq}. The latter is 
based on generating replica sets by introducing Gaussian fluctuations on the central PDF set with a variance 
determined by the PDF uncertainty given by the eigenvectors. Both reweighting methods are implemented in \fitter.
Note that the precise form of the weights used by both methods has recently been questioned \cite{Sato:2013ika,Paukkunen:2014zia}.

The Bayesian Reweighting technique relies on the fact that MC replicas of a PDF set give 
a representation of the probability distribution in the space of PDFs. In particular, the PDFs are represented 
as ensembles of $N_{\rm rep}$ equiprobable ({\em i.e.} having weights equal to unity) replicas, $\{f\}$. 
The central value for a given observable, $\mathcal{O}(\{f\})$, is computed as the average of the 
predictions obtained from the ensemble as
\begin{equation}
\langle\mathcal{O}(\{f\})\rangle =  \frac{1}{N_{\mathrm{rep}}} \sum_{k=1}^{N_{\mathrm{rep}}} \mathcal{O}(f^{k}),
\end{equation}
and the uncertainty as the standard deviation of the sample.

Upon inclusion of new data the prior probability distribution, given by the original PDF set, is modified according
to Bayes Theorem such that the weight of each replica, $w_k$, is updated according to
\begin{equation}
 w_k = \frac{(\chi^2_k)^{\frac{1}{2} (N_{\mathrm{data}}-1) } e^{-\frac{1}{2}\chi^2_k}}
          { \frac{1}{N_{\mathrm{rep}}} \sum^{N_{\mathrm{rep}}}_{k=1}(\chi^2_k)^{\frac{1}{2}(N_{\mathrm{data}}-1)} e^{-\frac{1}{2}\chi^2_k}  },
\end{equation}
where $N_{\mathrm{data}}$ is the number of new data points, $k$ denotes the specific replica for which the 
weight is calculated and $\chi^2_k$ is the $\chi^2$ of the new data obtained using the $k$-th PDF replica.
Given a PDF set and a corresponding set of weights, which describes the impact of the
inclusion of new data, the prediction for a given observable after inclusion of the new data can be computed as the {\em weighted} average,
\begin{equation}
\langle\mathcal{O}(\{f\})\rangle =  \frac{1}{N_{\mathrm{rep}}} \sum_{k=1}^{N_{\mathrm{rep}}} w_k \mathcal{O}(f^{k}).
\end{equation}

To simplify the use of a reweighted set, an unweighted set ({\em i.e.} a set of equiprobable replicas which incorporates 
the information contained in the weights) is generated according to the unweighting procedure described in~\cite{Ball:2011gg}. 
The number of effective replicas of a reweighted set is measured by its Shannon 
Entropy~\cite{Ball:2010gb}
\begin{equation}
\label{eq:shannon}
N_\mathrm{eff}\equiv 
\exp\left\{\frac{1}{N_\mathrm{rep}}\sum_{k=1}^{N_\mathrm{rep}}w_k\ln(N_\mathrm{rep}/w_k)\right\}\,,
\end{equation}
which corresponds to the size of a refitted equiprobable replica set containing the same amount of information. 
This number of effective replicas, $N_\mathrm{eff}$, gives an indicative measure of the optimal size of an 
unweighted replica set produced with the reweighting/unweighting procedure. No extra information is 
gained by producing a final unweighted set that has a number of replicas (significantly) larger than 
$N_\mathrm{eff}$.  If $N_\mathrm{eff}$ is much smaller than the original number of replicas the new data have great impact, however, it is unreliable to use the new reweighted set. In this case, instead, a full refit should be performed.


\section{Alternatives to DGLAP Formalism}
\label{sec:alternative}

QCD calculations based on the DGLAP~\cite{Gribov:1972ri,Gribov:1972rt,Lipatov:1974qm,
Dokshitzer:1977sg,Altarelli:1977zs} evolution equations are very successful in describing
all relevant hard scattering data in the perturbative region $Q^2 \gtrsim $ few $ \GeV^2$.
At small-$x$ ($x <$ 0.01) and small-$Q^2$ DGLAP dynamics may be modified
 by saturation and other (non-perturbative) higher-twist effects.
Different approaches alternative to the DGLAP formalism can be used to analyse DIS data in \fitter.
These include several dipole models and the use of 
transverse momentum dependent, or unintegrated PDFs (uPDFs).

\subsection{Dipole Models}

The dipole picture provides an alternative approach to proton-virtual photon
 scattering at low $x$ which can be applied to both inclusive and 
diffractive processes.
 In this approach, the virtual photon fluctuates into a $q\bar q$ (or $q\bar q g$) 
 dipole which interacts with the proton~\cite{NNZ:91,Mueller:1993rr}.  
The dipoles can be considered as quasi-stable quantum mechanical states, which have very long 
life time $\propto 1/m_p x\;$ and a size which is not changed by scattering with the proton.
The dynamics of the interaction are embedded in a dipole scattering amplitude.

Several dipole models, which assume different behaviours of the dipole-proton 
cross section, are implemented in \fitter:
the Golec-Biernat-W\"usthoff (GBW)
dipole saturation model~\cite{Golec-Biernat:1998js},
a modified GBW model which takes into account the effects of  
DGLAP evolution, termed the Bartels-Golec-Kowalski (BGK) dipole model~\cite{Bartels:2002cj}
and the colour glass condensate approach to the high parton density 
regime, named the Iancu-Itakura-Munier (IIM) dipole model~\cite{Iancu:2003ge}.

\paragraph{GBW model:} \rm
In the GBW model the dipole-proton cross section $\sigma_{\text{dip}}$ is given by
\begin{equation}
\label{eGBW}
   \sigma_{\text{dip}}(x,r^{2}) = \sigma_{0} \left(1 - \exp \left[-\frac{r^{2}}{4R_{0}^{2}(x)} \right]\right),
\end{equation}
where $r$ corresponds to the transverse separation between the quark and the antiquark, and $R_{0}^{2}$
 is an $x$-dependent scale parameter which represents the spacing of the gluons in the proton. 
$R_{0}^{2}$ takes the form, $R_0^2(x) = (x/x_0)^\lambda  1/ {\rm \GeV}^{2}$, and is called the saturation radius.
The cross-section normalisation $\sigma_0$, $x_0$, and $\lambda$ are parameters 
of the model fitted to the DIS data.
This model gives exact Bjorken scaling when the dipole size $r$ is small.
 
\paragraph{BGK model:} \rm
The BGK model is a modification of the GBW model assuming that the
spacing $R_0$ is inverse to the gluon distribution and taking
into account the DGLAP evolution of the latter.
The gluon distribution, parametrised at some starting scale by Eq.~\ref{eqn:pdf_std}, 
is evolved to larger scales using DGLAP evolution.

\paragraph{BGK model with valence quarks:} \rm
The dipole models are valid in the low-$x$ region only, where the valence quark contribution to the total proton momentum 
is 5\% to 15\% for $x$ from 0.0001 to 0.01 \cite{Collaboration:2010ry}.
The inclusive HERA measurements have a precision which is better than 2$\%$. 
Therefore, \fitter provides the option of taking into account the contribution of the valence quarks
%

\paragraph{IIM model:} \rm
The IIM model assumes an expression for the dipole cross section which is based on the 
Balitsky-Kovchegov equation~\cite{Balitsky:1995ub}. The explicit formula for $\sigma_{\text{dip}}$ 
can be found in~\cite{Iancu:2003ge}. 
The alternative scale parameter $\tilde{R}$, $x_{0}$ and $\lambda$ are fitted parameters of the model.

\subsection{Transverse Momentum Dependent PDFs}

\def\kt{\ensuremath{k_t}}
\def\pt{\ensuremath{p_t}}

QCD calculations of multiple-scale processes  and complex final-states
can necessitate the use of transverse-momentum dependent (TMD)~\cite{Collins:2011zzd}, or 
unintegrated parton distribution and parton decay 
functions~\cite{Aybat:2011zv,Buffing:2013eka,Buffing:2013kca,Buffing:2012sz,Mulders:2008tf,Jadach:2009gm,Hautmann:2009zzb,Hautmann:2012pf,Hautmann:2007gw}.   
TMD factorisation has been proven recently \cite{Collins:2011zzd} for inclusive DIS. TMD factorisation has also been proven in the high-energy (small-$x$) limit \cite{Catani:1990xk,Collins:1991ty,Hautmann:2002tu} for 
particular hadron-hadron scattering processes, like heavy flavour, vector boson and Higgs production.
  
In the framework of high-energy factorisation~\cite{Catani:1990xk,Catani:1990eg,Catani:1993ww} 
the DIS cross section can be written as a convolution in 
both longitudinal and transverse momenta of the TMD parton distribution function 
${\cal A}\left(x,\kt,\mu_{F}^2\right)$    
 with the off-shell parton scattering matrix elements as follows 
\begin{equation}
 \sigma_j (x, Q^2) = \int_x^1  
d z \int d^2k_t \ 
\hat{\sigma}_j(x,Q^2,{z},k_t) \ 
 {\cal  A}\left( {z} ,\kt, \mu_{F}^2 \right), 
\label{kt-factorisation}
\end{equation}
where  the DIS cross sections 
$\sigma_j$($j= 2 , L$) 
are related to the  structure functions $F_2$ and $F_L$ 
by 
$\sigma_j = 4 \pi^2 F_j / Q^2$, and   
the hard-scattering kernels ${\hat \sigma}_j$ of Eq.~\ref{kt-factorisation}
are $k_t$-dependent. 

The factorisation formula in Eq. \ref{kt-factorisation}
allows resummation of   
logarithmically enhanced small-$ x  $ contributions  
 to all orders in perturbation theory,  
both in the  hard 
scattering coefficients and 
in  the parton evolution, fully taking into account the 
dependence on the factorisation scale $\mu_F$ and on the 
factorisation scheme~\cite{Catani:1994sq,Catani:1993rn}.

Phenomenological applications of this approach require 
matching of small-$ x  $ contributions with  finite-$x$ contributions.  To this end, 
the evolution  of the 
transverse momentum dependent gluon density 
${\cal  A} $ is obtained  by combining    the resummation of  small-$x$ logarithmic 
corrections~\cite{Lipatov:1996ts,Fadin:1975cb,Balitsky:1978ic}   with   medium-$x$  and large-$x$ 
contributions to parton  splitting~\cite{Gribov:1972ri,Altarelli:1977zs,Dokshitzer:1977sg} according to the 
CCFM  evolution equation~ \cite{\CCFM}. Sea quark
contributions \cite{Hautmann:2012sh} are not yet
included at transverse momentum dependent level. 

The cross section $\sigma_j$, ($j= 2 , L$) is calculated in a FFN   scheme,  using 
the boson-gluon fusion process ($\gamma^* g^* \to q \bar{q}$). The masses of the 
quarks are explicitly included as  parameters of the model.
In addition to $\gamma^* g^* \to q\bar{q}$,  the contribution from valence quarks is included 
via $\gamma^* q \to q$   by using  a CCFM evolution of 
valence quarks~\cite{Deak:2010gk,Deak:2011ga,Hautmann:2013tba}. 

  
 

\paragraph{CCFM Grid Techniques:} \rm

The CCFM evolution cannot be written easily in an analytic closed form. For this 
reason, a MC method is employed, which is, however, time-consuming, and thus
cannot be used directly in a fit program. 

Following the  convolution method introduced in~\cite{Jung:2012hy,Hautmann:2013tba}, the 
kernel $ \tilde {\cal A}\left(x'',\kt,\Pmax\right) $ is determined from the MC  solution of the CCFM evolution equation, 
and then folded with a non-perturbative starting distribution ${\cal A}_0 (x)$
{ 
\begin{eqnarray}
x {\cal A}(x,\kt,\Pmax) &= &x\int dx' \int dx'' {\cal A}_0 (x') \tilde{\cal A}\left(x'',\kt,\Pmax\right) 
 \delta(x' 
x'' - x) 
\nonumber  
\\
& = & \int dx' {{\cal A}_0 (x') }  
\frac{x}{x'} \ { \tilde{\cal A}\left(\frac{x}{x'},\kt,\Pmax\right) }, 
\end{eqnarray}
}
where $\kt$ denotes the transverse momentum of the propagator gluon and $\Pmax$ is the 
evolution variable.

The kernel $\tilde{\cal A}$ incorporates all of  the dynamics of the evolution.  
It is defined on a grid of $50\otimes50\otimes50$ bins in $ x, \kt, \Pmax$.  
The binning in the grid is logarithmic, except for the longitudinal variable 
$x$ for which 40 bins in logarithmic 
spacing below 0.1, and 10 bins in linear spacing above 0.1 are used.

Calculation of the cross section according to Eq.~\ref{kt-factorisation} involves a time-consuming
multidimensional MC integration, which suffers from numerical fluctuations.  
This cannot be employed directly in a fit procedure. Instead the following equation is applied:
\begin{eqnarray}
\sigma(x,Q^2) & = & \int_x^1 d x_g {\cal A}(x_g,\kt,\Pmax) \hat{ \sigma}(x,x_g,Q^2) 
\nonumber\\
  & = & \int_x^1 dx' {\cal A}_0 (x')  \tilde{ \sigma}(x/x',Q^2),
    \label{final-convolution}
 \end{eqnarray}
where first $ \tilde{ \sigma}(x',Q^2)$ is calculated numerically with a MC integration 
on a grid in $x$ for the values of $Q^2$ used in the fit. Then the last step in Eq.~\ref{final-convolution}  
is performed with a fast numerical Gauss integration, which can be used directly in the fit.

\paragraph{Functional Forms for TMD parametrisation:} \rm

For the starting distribution ${\cal A}_0$, at the starting scale $Q_0^2$, 
the following form is used:
\begin{eqnarray}
x{\cal A}_0(x,\kt) &=& N x^{-B}(1 -x)^{C}\left(1 -D x 
+ E \sqrt{x}   \right) \nonumber\\
   &\times &\exp[ - k_t^2 / \sigma^2 ], 
\label{a0-5par}
\end{eqnarray}
where $ \sigma^2  =  Q_0^2 / 2 $ and $N, B, C, D, E$ are free parameters.
Valence quarks are treated using the method of Ref.~\cite{Deak:2010gk} as described 
in Ref. \cite{Hautmann:2013tba} with a starting distribution taken from any collinear PDF
and imposition of the flavour sum rule at every scale $p$.
\\
The TMD parton densities can be plotted either with \fitter tools 
or with \tt TMDplotter\rm~\cite{Hautmann:2014kza}.


\section{\fitter Code Organisation}
\label{sec:organisation}

\begin{figure}[!ht]
   \centering
   \includegraphics[width=8.65cm]{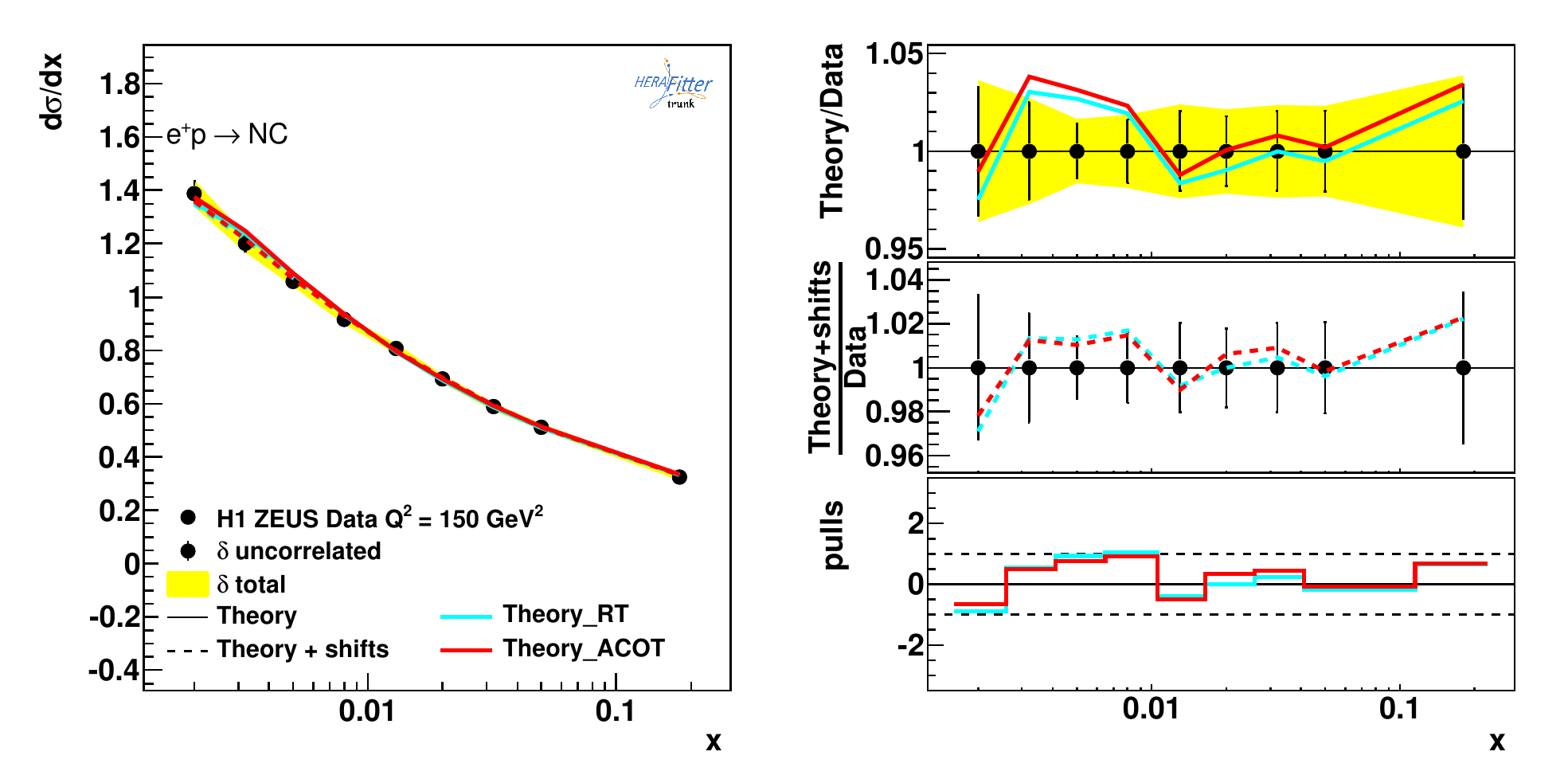}
   \caption{An illustration of the consistency of HERA measurements~\cite{h1zeus:2009wt} and the theory predictions, 
       obtained in \fitter with the default drawing tool.} 
 \label{fig:data}
\end{figure}

\fitter is an open source code under the GNU general public licence. It can be downloaded from a dedicated 
webpage \cite{herafitter:page}
together with its supporting documentation and 
\emph{fast grid} theory files (described in Sec. \ref{sec:techniques}) associated with data files.
The source code contains all the relevant information to perform QCD fits with HERA DIS data as a default 
set. \footnote{Default settings in \fitter are tuned to reproduce the central HERAPDF1.0 set.} 
The execution time depends on the fitting options and varies from 10 minutes 
(using ``FAST'' techniques as described in Sec. \ref{sec:techniques}) to several hours when 
full uncertainties are estimated. The \fitter code is a combination of \tt C++ \rm and \tt Fortran 77\rm \ libraries with minimal 
dependencies, i.e. for the default fitting options no external dependencies are required except the \qcdnum evolution program \cite{qcdnum}.
The \tt ROOT \rm  libraries are only required for the drawing tools and when invoking \applgrid.  
Drawing tools built into \fitter provide a qualitative and quantitative assessment of the results.
Fig.~\ref{fig:data} shows an illustration of a comparison between the inclusive NC data from HERA I
with the predictions based on HERAPDF1.0 PDFs.
The consistency of the measurements and the theory can be expressed by pulls, defined as the difference between data and theory divided by the uncorrelated error of the data. 
In each kinematic bin of the measurement, pulls are provided in units of standard deviations.  
The pulls are also illustrated in Fig.~\ref{fig:data}.

In \fitter there are also available cache options for fast retrieval, fast evolution kernels, and the OpenMP (Open Multi-Processing) 
interface which allows parallel applications of the GM-VFNS theory predictions in DIS.

\section{Applications of \fitter}
\label{sec:examples}
The \fitter program has been used in a number of experimental and theoretical analyses. 
This list includes several LHC analyses of SM processes, namely
inclusive Drell-Yan and $W$and $Z$ production~\cite{atlas:strange,cms:strange,atlas:hm,Aad:2014qja,Aad:2014xca}, 
inclusive jet production \cite{atlas:jets,Khachatryan:2014waa}, and inclusive photon production \cite{atlas:photons}.
The results of QCD analyses using \fitter were also
published by HERA experiments for inclusive \cite{h1zeus:2009wt,h1:2012kk} and
 heavy flavour production measurements \cite{h1zeus:charm, Abramowicz:2014zub}.
The following phenomenological studies have been performed with \fitter:
a determination of the transverse momentum dependent gluon distribution using precision HERA data \cite{Hautmann:2013tba}, 
an analysis of HERA data within a dipole model \cite{Luszczak:2013rxa},
the study of the low-x uncertainties in PDFs determined from the HERA data using 
different parametrisations \cite{Chebyshev}. It is also planned to use \fitter for studying  the impact of QED radiative corrections on PDFs \cite{Sadykov:2014aua}.
A recent study based on a set of PDFs determined with \fitter and addressing 
the correlated uncertainties between different orders has been published in \cite{hfcorrpaper}. 
An application of the TMDs obtained with \fitter
to $W$ production at the LHC can be found in \cite{Dooling:2014kia}.

The \fitter framework has been used to produce PDF grids from QCD analyses performed at 
HERA \cite{h1zeus:2009wt,hera:grids} and at the LHC \cite{atlas:grids}, using 
measurements from ATLAS~\cite{atlas:strange,atlas:jets}. These PDFs can be used to study predictions for SM 
or beyond SM processes. Furthermore, \fitter provides the possibility to perform various benchmarking exercises
\cite{Butterworth:2014efa} and impact studies for possible future colliders
as demonstrated by QCD studies at the LHeC~\cite{lhec:studies}.

\section{Summary}
\label{sec:outlook}

\label{sec:summary}
\fitter is the first open-source code designed for studies of the structure of the proton.
It provides a unique and flexible framework with a wide variety of QCD tools to 
facilitate analyses of the experimental data and theoretical calculations. 

The \fitter code, in version $1.1.0$, has sufficient options to reproduce the majority of the different 
theoretical choices made in MSTW, CTEQ and ABM fits. This will potentially make it a  
valuable tool for benchmarking and understanding differences between PDF fits. 
Such a study would however need to consider a range of further questions, such as the choices of
data sets, treatments of uncertainties, input parameter values, $\chi^2$ definitions, nuclear corrections, etc. 
\\
The further progress of \fitter will be driven by the latest QCD advances in theoretical calculations 
and in the precision of experimental data.



%
%

\begin{acknowledgements}
\fitter developers team acknowledges the kind hospitality of DESY 
and funding by the Helmholtz Alliance "Physics at the Terascale" of the Helmholtz Association.
We are grateful to the DESY IT department for their support of the \fitter  developers. We thank the H1 and ZEUS collaborations for the support in the initial stage of the project.
Additional support was received from the BMBF-JINR cooperation program,
the Heisenberg-Landau program, the RFBR grant 12-02-91526-CERN~a, 
the Polish NSC project DEC-2011/03/B/ST2/00220 and 
a dedicated funding of the Initiative and Networking Fond of Helmholtz Association SO-072. 
We also acknowledge Nathan Hartland with Luigi Del Debbio for contributing to the 
implementation of the Bayesian Reweighting technique and
would like to thank R. Thorne for fruitful discussions.
\end{acknowledgements}


\bibliography{herafitter-epjc.bib}



\end{document}